\documentclass[reprint,twocolumn,showpacs,superscriptaddress,prb,floatfix]{revtex4}
\usepackage{dcolumn}
\usepackage{natbib}

\usepackage{bm}        
\usepackage{amssymb}   
\usepackage{color,soul} 

\usepackage[pdftex]{graphicx}
\usepackage[pdftex]{epsfig}
\usepackage{epstopdf}
\usepackage{dcolumn}
\usepackage{bm}

\newcommand{\units}[1]{\ensuremath{\mathrm{#1}}}
\newcommand{\amount}[2]{\ensuremath{#1\:\units{#2}}}

\newcommand{\Soo}[0]{S$_{1,1}$\ } 
\newcommand{\Stz}[0]{S$_{2,0}$\ } 
\newcommand{\Too}[0]{T$_{1,1}$\ } 
\newcommand{\Ttz}[0]{T$_{2,0}$\ } 

\newcommand{\VSD}{\ensuremath{V_{\mathrm{SD}}}}

\begin{document}

\title{Pauli Spin Blockade and Lifetime-Enhanced Transport in a
  Si/SiGe double quantum dot}

\author{C. B. Simmons} 
\affiliation{University of Wisconsin-Madison, Madison, Wisconsin 53706, USA}

\author{Teck Seng Koh} 
\affiliation{University of Wisconsin-Madison, Madison, Wisconsin 53706, USA}

\author{Nakul Shaji} 
\affiliation{University of Wisconsin-Madison, Madison, Wisconsin 53706, USA}

\author{Madhu Thalakulam} 
\affiliation{University of Wisconsin-Madison, Madison, Wisconsin 53706, USA}

\author{L. J. Klein} 
\affiliation{University of Wisconsin-Madison, Madison, Wisconsin 53706, USA}

\author{Hua Qin} 
\affiliation{University of Wisconsin-Madison, Madison, Wisconsin 53706, USA}

\author{H. Luo} 
\affiliation{University of Wisconsin-Madison, Madison, Wisconsin 53706, USA}

\author{D. E. Savage} 
\affiliation{University of Wisconsin-Madison, Madison, Wisconsin 53706, USA}

\author{M. G. Lagally} 
\affiliation{University of Wisconsin-Madison, Madison, Wisconsin 53706, USA}

\author{A. J. Rimberg} 
\affiliation{Department of Physics and Astronomy, Dartmouth College, Hanover, New Hampshire 03755, USA}

\author{Robert Joynt} 
\affiliation{University of Wisconsin-Madison, Madison, Wisconsin 53706, USA}

\author{Robert Blick} 
\affiliation{University of Wisconsin-Madison, Madison, Wisconsin 53706, USA}

\author{Mark Friesen} 
\affiliation{University of Wisconsin-Madison, Madison, Wisconsin 53706, USA}

\author{S. N. Coppersmith}
 \affiliation{University of Wisconsin-Madison, Madison, Wisconsin 53706, USA}

\author{M. A. Eriksson} 
\affiliation{University of Wisconsin-Madison, Madison, Wisconsin 53706, USA}

\begin{abstract}
We analyze electron transport data through a Si/SiGe double quantum dot in terms of spin blockade and lifetime-enhanced transport (LET), which is transport through excited states that is enabled by long spin relaxation times.  We present a series of low-bias voltage measurements showing the sudden appearance of a strong tail of current that we argue is an unambiguous signature of LET appearing when the bias voltage becomes greater than the singlet-triplet splitting for the (2,0) electron state.  We present eight independent data sets, four in the forward bias (spin-blockade) regime and four in the reverse bias (lifetime-enhanced transport) regime, and show that all eight data sets can be fit to one consistent set of parameters.  We also perform a detailed analysis of the reverse bias (LET) regime, using transport rate equations that include both singlet and triplet transport channels.  The model also includes the energy dependent tunneling of electrons across the quantum barriers, and resonant and inelastic tunneling effects.  In this way, we obtain excellent fits to the experimental data, and we obtain quantitative estimates for the tunneling rates and transport currents throughout the reverse bias regime.  We provide a physical understanding of the different blockade regimes and present detailed predictions for the conditions under which LET may be observed.
\end{abstract}

\pacs{73.23.Hk, 73.21.La, 73.63.Kv, 85.35.Gv, 81.05.Cy, 03.67.Lx}

\maketitle

\section{\label{intro}Introduction}
Spins in quantum dots are good candidates for qubits, due to fast and reliable electrostatic gating, long coherence times, and well known methods for large-scale fabrication.\cite{Loss:1998p120,Kane:1998p133,Divincenzo:2000p1642,Vrijen:2000p1643,Friesen:2003p121301}  Spin coherence times can be particularly long in silicon devices,\cite{Feher:1959p1245,Tyryshkin:2003p193207,Tyryshkin:2006p36,Hayes:2009preprint,Xiao:2010p1876,Morello:2010p1933} because the naturally abundant $^{28}$Si isotope has nuclear spin zero, and because Si has a relatively small spin-orbit coupling, because of its relatively small atomic mass.  Isotopic purification could produce devices with excellent qualities for quantum computing.\cite{Tahan:2002p035314,Assali:2010preprint}

Recent progress in GaAs quantum dots has enabled the manipulation of exchange coupling in a two-electron double dot,\cite{Petta:2005p2180} and has led to single-shot readout of one- \cite{Elzerman:2004p431} and two-electron \cite{Hanson:2005p719} spin states in a single dot.  The latter experiment makes use of energy dependent tunneling to provide high visibility in the measurement.  The simplest explanation of this effect is that a larger tunnel barrier causes a slower tunneling rate.  The effect is of fundamental interest for quantum phenomena, and can lead to very precise measurement techniques in the context of quantum information.  Energy dependent tunneling effects in quantum dots have also been studied in several other recent experiments.\cite{Petta:2005p161301,Ono:2005p1313,Johnson:2005p925,Johnson:2005p483,Maclean:2007p1499,Amasha:2008p1500}
Silicon-based devices should exhibit a strong energy dependence in tunneling, since the effective mass in silicon, on which the tunneling rate depends exponentially,\cite{DaviesBook} is relatively large.

Semiconductor double quantum dots are tunable structures that enable
the coupling of two small regions of bound electrons to each other
and, often, to two leads, enabling measurement of an electron
transport current through the system.\cite{VanDerWiel:2002p1382} Double quantum dots can display an effect
known as Pauli spin blockade,\cite{Rokhinson:1998p1452,Huttel:2003p1958,Ono:2005p1313} where current flow proceeds
in a cycle that first loads either a two-electron singlet state with
one electron in each dot, the \Soo state, or a two-electron triplet
\Too state that is nearly degenerate with the singlet. For the cycle
to proceed without blockade in either case, both the singlet \Stz and the
triplet \Ttz states, where both electrons are on the left dot, must be lower in energy than their corresponding
(1,1) states.  There are regions in gate-voltage space where this is
not true, and the striking absence of current that arises in such regions is known as
spin blockade. Spin blockade makes double dots extremely useful for
quantum dot spin qubits, because it provides a robust means to perform
spin readout.\cite{Levy:2002p1446, Taylor:2005p482, Petta:2005p2180, Johnson:2005p925,
  Nowack:2007p1430, PioroLadriere:2008p776}

Spin blockade has been reported in silicon quantum dots formed using
both Si metal-oxide-semiconductor structures \cite{Liu:2008p073310}
and Si/SiGe heterostructures.\cite{Shaji:2008p540}  In both cases,
the spin blockade results displayed many similarities with previous
observations of spin blockade, most of which have been made in
GaAs/AlGaAs-based double quantum dots. In Ref.~\onlinecite{Shaji:2008p540},
we also reported measurements of current flow for the opposite
voltage bias.  In this regime, we observed unusual patterns in two-dimensional
maps of the current as a function of a pair of gate voltages. The most
striking observation was a strong `tail' of current, which appeared in a supposedly 
blockaded portion of the current map.  This behavior was attributed to the combined
effects of slow triplet-singlet relaxation and strong energy dependent
tunneling, the two of which together enable current to flow through
long-lived excited states.  For this reason, the phenomena were
labeled `lifetime-enhanced transport,' or LET.  
Recently, LET behavior has also been observed in transport through individual donors in silicon.\cite{Lansbergen:2010preprint}

In this paper, we present a large quantity of additional data
and analyze these data in detail, showing that they
can all be fit using one consistent set of parameters.  We
analyze eight two-dimensional plots of current as a function of a pair
of gate voltages, four different biases in each direction of current
flow through the double dot.  Spin blockade is observed in the
direction of current flow in which it is expected.  For the opposite voltage bias, the unusual
current patterns associated with LET are observed, in agreement with our previous results.\cite{Shaji:2008p540}
To test the explanation that the `tail' of current arises because of LET, we fit 
all eight sets of data in parallel, determining optimum values for
the three slopes of the edges of the bias triangles, the lengths of the
sides of the bias triangles, and the positions of those triangles. The
results are shown to be consistent with both spin blockade and LET. In particular, it is shown that
we cannot obtain consistent results for the length of the triangles if the current tails are 
included in the triangles.  The fitting
and the subsequent delineation of the bias triangles also enables us to improve our
measurements of the singlet-triplet splittings in both the (2,0) state, \textcolor{black}{corresponding to two electrons in the left dot,} and the (1,1) state, \textcolor{black}{corresponding to one electron in each dot.}

Building on accurate fitting of the bias triangles,
we investigate details of the strong dependence of
the transport current on the gate voltages.  Within the triangles, we obtain
consistent and quantitative fits to the data by explicitly incorporating strong energy-dependent tunneling as well as tunneling through both singlet and triplet channels.  Both coherent\cite{Nazarov:1993p57,Stoof:1996p1050} and incoherent\cite{Fujisawa:1998p932} processes contribute strongly to the energy dependence of the electron tunneling rates, and we develop a general model for transport in a double dot, including both inelastic and resonant effects, using the master equation approach.  We show that this model can be used to perform a quantitative fit of the transport data and that relevant parameters can be extracted from the data.

Our success in interpreting a large body of experimental data with a single consistent
set of fitting parameters is strong evidence in support of the interpretation of the
current tail in terms of
transport through the triplet channel, 
as first described in Ref.~\onlinecite{Shaji:2008p540}.

The paper is organized as follows.
Sec.~\ref{sec:simple_demonstration} presents a simple argument
for the existence of the LET tail based on data at small bias voltages and involves
an absolute minimum of data analysis.
Sec.~\ref{sec:exp} presents our experimental procedures.
The methods used to fit the bias triangles are presented in Sec.~\ref{sec:fitting},
and the procedure used to position and scale the bias triangles are
presented in Sec.~\ref{sec:positioning}.
The phenomenon of energy-dependent
tunneling is prominent in the data, and a theoretical model that describes this effect as it appears in the data is
presented in Sec.~{\ref{sec:theory}}.  In Sec.~{\ref{sec:data_energy}}, results from the model are compared to the experimental data, and important phenomenological parameters are reported.  The paper concludes with a discussion in Sec.~{\ref{sec:discussion}}.

\begin{figure}
\includegraphics[width=3.4in]{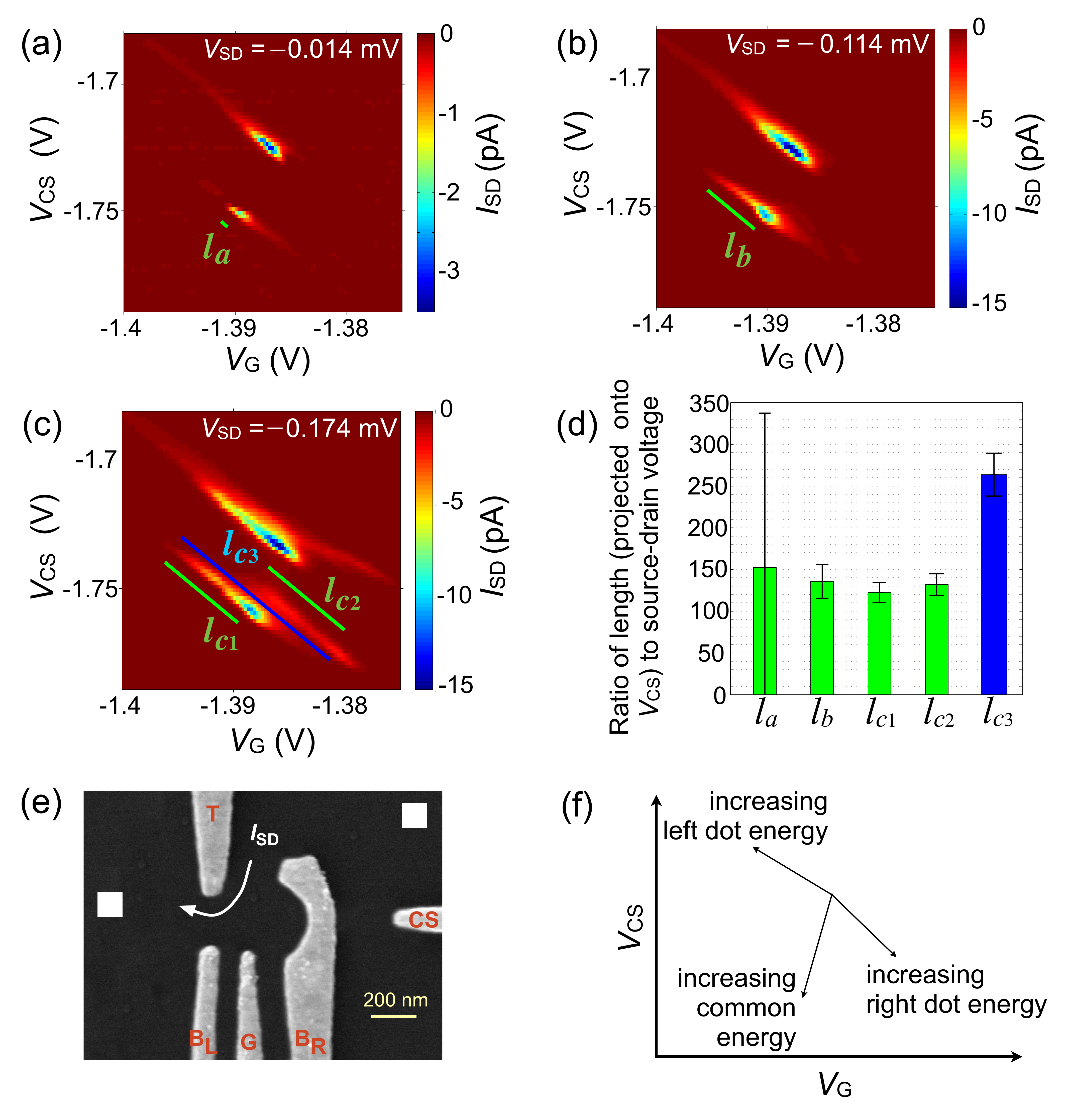}
\caption{\label{fig:simple}The transport current through a double quantum
  dot for a device with gate arrangement shown in the inset to (a)
  at three different source-drain biases \VSD: (a)
  \amount{-0.014}{mV}, (b) \amount{-0.114}{mV}, (c)
  \amount{-0.174}{mV}. Panel (c) shows the sudden appearance of a current tail.  (d) Lengths, $l_a$, $l_b$ and $l_{c_1}$ in panels (a) to (c) are measured from the peak of the current to the upper left tip of a 1 pA contour around the peak.  $l_{c_2}$ is the length of the tail from its peak at the lower right to the tip of a 1 pA contour on the upper left. $l_{c_3}$ is the length that would be extracted if the `tail' were part of the conventional bias triangle. It is measured from the peak of the current in the tail to the tip of the upper left-most 1 pA contour; i.e., it is essentially the sum of $l_{c_1}$ and $l_{c_2}$. The graph in (d) shows the ratio of the lengths, as measured by their projection onto the left-hand gate voltage axis, divided by \VSD.  The four lengths $l_a$, $l_b$, $l_{c_1}$, and $l_{c_2}$ are consistent with each other and with the description of the `tail' in terms of LET.
The blue length $l_{c_3}$ is clearly too long, and thus the description of the physics in terms of LET and the concept of a current `tail' extending out of the conventional bias triangle are necessary to understand the data in panels (a)-(c).
\textcolor{black}{(e) SEM image of the gate pattern with the gates labeled
  in red.  The gates were tuned so that the device contained a double quantum dot.\cite{Shaji:2008p540}
  (f) Schematic diagram of the energy axes for the double dot.} }
\end{figure}

\section{Simple Demonstration of LET}
\label{sec:simple_demonstration}

The key observable feature of LET is a strong tail of current
protruding beyond the end of the base of the conventional bias
triangle. Careful fitting of the triangles to many sets of data obtained with a variety of source-drain bias voltages, as we do below, is a good way to test for the existence
of this tail.  However, before embarking on such a detailed analysis, we first provide a simple  
argument for our LET interpretation of the data.

Figure~\ref{fig:simple}(a-c) show plots of the current through a double
quantum dot as a function of two gate voltages.  The sample is described in Sec.~\ref{sec:exp} below, and the sign of the voltage
bias is opposite to that in which spin blockade would be
observed.  Fig.~\ref{fig:simple}(a) is obtained at very low bias, such that
current flows only 
very near the triple points.\cite{VanDerWiel:2002p1382} In
particular, the bias voltage is smaller than the singlet-triplet
splitting. Fig.~\ref{fig:simple}(b) shows data at a slightly larger
voltage. As expected, the region of current flow is correspondingly larger, and it has the same
overall shape.  Fig.~\ref{fig:simple}(c) is acquired at a bias voltage
slightly higher still.  Again, the primary region of current flow is
correspondingly larger.  However, a new
feature suddenly appears in the data: a tail of current that extends to the
lower right. This tail completely changes the shape of the current
pattern.

To be quantitative, we consider the lengths of the various current
features visible in Fig.~\ref{fig:simple}(a)-(c).  The bias triangles, and therefore the lengths of these current features, should scale linearly with the applied bias voltage.  Panel (d) shows in green the ratio of the lengths of the
 main current features identified in panels (a)-(c), projected onto the left-hand voltage axis, to the applied bias voltage \VSD.  
These lengths were measured, as indicated, from the point where the current peaks at the lower right, to the extremum at the upper left on a 1 pA contour.
The LET interpretation predicts that the ratios of lengths $l_a$, $l_b$, and $l_{c_1}$ to $V_\mathrm{SD}$ should be equal, as shown in panel~(d).  In particular, $l_b$ and $l_{c_1}$, which have small error bars, are nearly identical.
Further, the length of the tail is the same as the length of the main current region in
Fig.~\ref{fig:simple}(c): that is, $l_{c1}=l_{c2}$.  This is also consistent with the LET interpretation presented below and in Ref.~\onlinecite{Shaji:2008p540}. In contrast, if the base of the triangle were located at the end of
the tail in Fig.~\ref{fig:simple}(c), the length of the region of
current flow would be as is shown in blue and labelled $l_{c3}$.  As is
clear from Fig.~\ref{fig:simple}(d), such a length is incompatible with length $l_b$ from
Fig.~\ref{fig:simple}(b).  Note that the length $l_a$ has a large error bar, because the source-drain voltage \VSD\ is so small that the uncertainty in that quantity is larger than its value; panel (a) is included in this discussion to emphasize that current is indeed visible at very low bias voltage, indicating that current flows at the triple points themselves.  If the base of the bias triangle were located at the end of the tail in Fig.~\ref{fig:simple}(c), no current would be observed at the triple point.  In such an interpretation, the triple point would be located directly below the tail, and this would be incompatible with the observation of current at the triple point in Fig.~\ref{fig:simple}(a).

These simple arguments make clear that the LET tail does indeed protrude
beyond the base of the bias triangle.  The rest of this paper presents
an analysis of a large quantity of data, all of which is analyzed
together and self-consistently.  The results of our analysis provide a complete and
quantitative understanding of the data in terms of spin blockade and LET.

\section{Experiment}
\label{sec:exp}
The data we discuss here were acquired from a double quantum dot formed
in a top-gated Si/Si$_{0.7}$Ge$_{0.3}$ heterostructure.\cite{Slinker:2005p246, Berer:2006p536, Klein:2007p033103} The
quantum well was nominally \amount{12}{nm} thick, and it contained
a two-dimensional electron gas of density \amount{n=4\times
  10^{11}}{cm^{-2}} and mobility \amount{40,000}{cm^{2}/Vs}. The gate
design for this device,
\textcolor{black}{reproduced from Ref.~\onlinecite{Shaji:2008p540} as Fig.~1(e),}
has a single plunger gate.  For the data presented here, a gate originally
intended for charge sensing was pressed into service to tune the dot
occupation, providing the second axis for manipulation of the double dot in
gate-voltage space.\cite{Shaji:2008p540}
\textcolor{black}{Here we focus on the region in gate-voltage
space where the device exhibits the behavior
of a double-dot; this region occurs between a regime in which the device
acts as a single dot and a region in which no measurable current flows
through the device.\cite{Shaji:2008p540}}
Unlike more recent work,\cite{Simmons:2007p213103,Thalakulam:2010p183104}
the absolute number of
electrons in the dots is not known, and all references to the number
of electrons refers to the valence number; there could be a closed
shell underneath the valence electrons, and the existence of that
shell would not be apparent in the data.

\begin{figure*}
\includegraphics[width=7in]{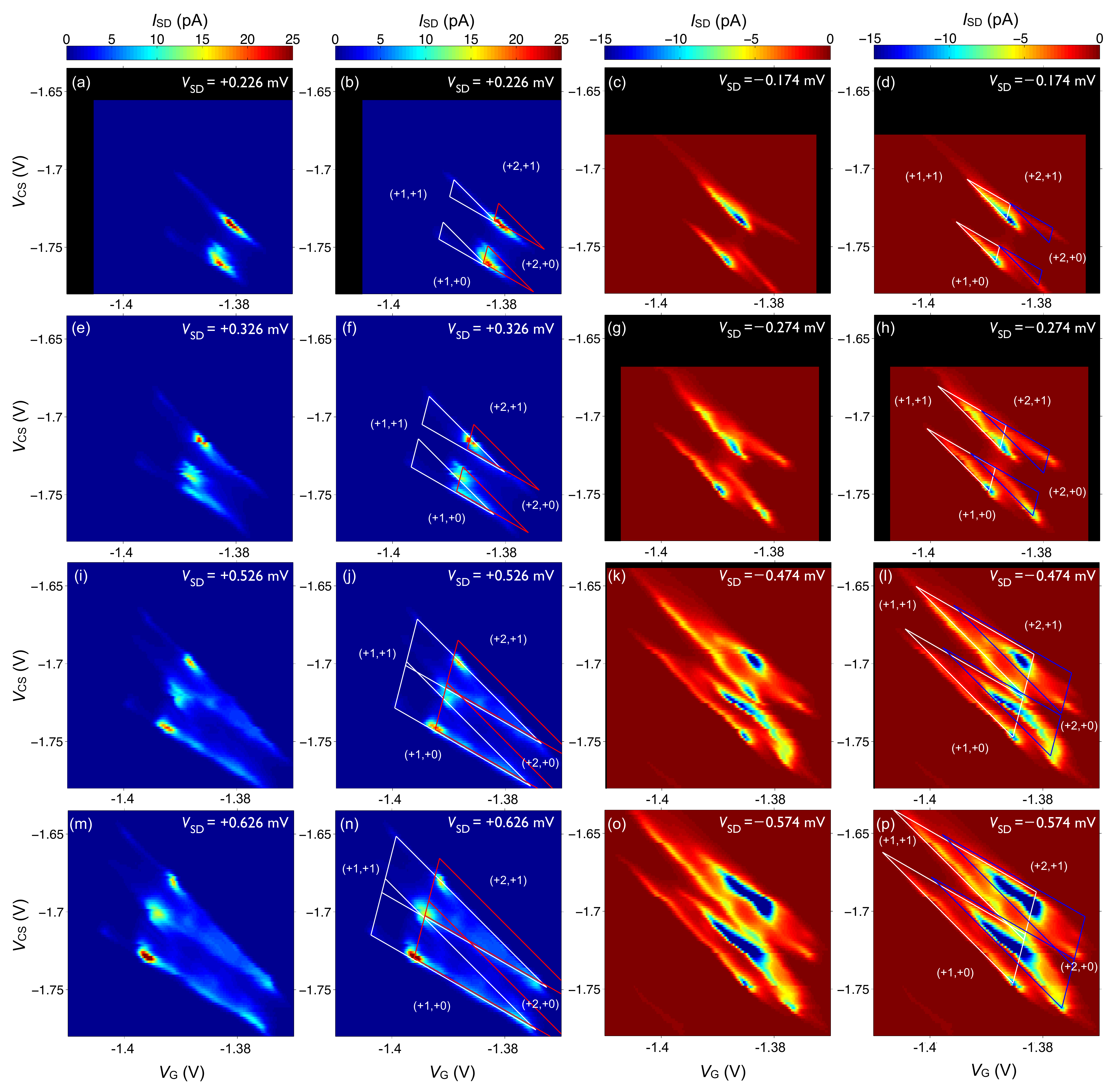}
\caption{\label{fig:all-data}The current through the double quantum dot as
  a function of the gate voltages $V_{\mathrm{G}}$ and
  $V_{\mathrm{CS}}$. 
  Panels (a)--(p) correspond to eight different
  bias voltages \VSD~as labelled. Column one shows data in the spin
  blockade regime (forward bias).  Column three shows the data in the
  LET regime (reverse bias).  Columns two and four show the same data
  as columns one and three, with the calculated bias triangle boundaries superimposed on the data, as explained in the main text.  The data in panels (a) and
  (g) have been presented previously in Ref.~\onlinecite{Shaji:2008p540}.}
\end{figure*}

Figure~\ref{fig:all-data} shows the current $I$ through the double quantum
dot as a function of gate voltages $V_{\mathrm{G}}$ and
$V_{\mathrm{CS}}$.  Each of the panels in Fig.~\ref{fig:all-data} contains
two features that are similar to each other. These features are
conventionally called the `electron' and `hole' triangles,\cite{VanDerWiel:2002p1382} and we adopt this language here. The
`electron' system is called so because it can be described in terms
of the electron occupations (1,1), (2,0), and (1,0).  The `hole' system can be described
in terms of the electron occupations (2,1), (2,0), and (1,1).  However,
these states result in energy level diagrams that are more complicated
than those of the electron triangle.  A complementary description of transport in terms of
the hole states (0,1)$_h$, (0,2)$_h$, and (1,1)$_h$ allows us to draw a
set of energy level diagrams that are analogous to the diagrams for the `electron' system.  This hole picture has proven useful in some situations.\cite{VanDerWiel:2002p1382}  However, it is difficult to include excited states of the
zero-hole state in a simple way. Thus, we will stick with the three-electron diagrams here,
in spite of their complexity.\cite{Koh:2010preprint} Detailed plots of the chemical potentials relevant for modeling of 
transport with both positive and negative voltage bias are shown in 
the Appendix 
in Fig.~\ref{chemical-potentials}.  We emphasize that, although for clarity and connection with the existing literature we retain the electron and hole terminology, the chemical potential diagrams for the latter in Fig.~\ref{chemical-potentials} actually describe three-electron states, not hole states.  In this paper, we refer to chemical potentials for electrons only, and never holes.

\begin{figure}
\includegraphics[width=3.4in]{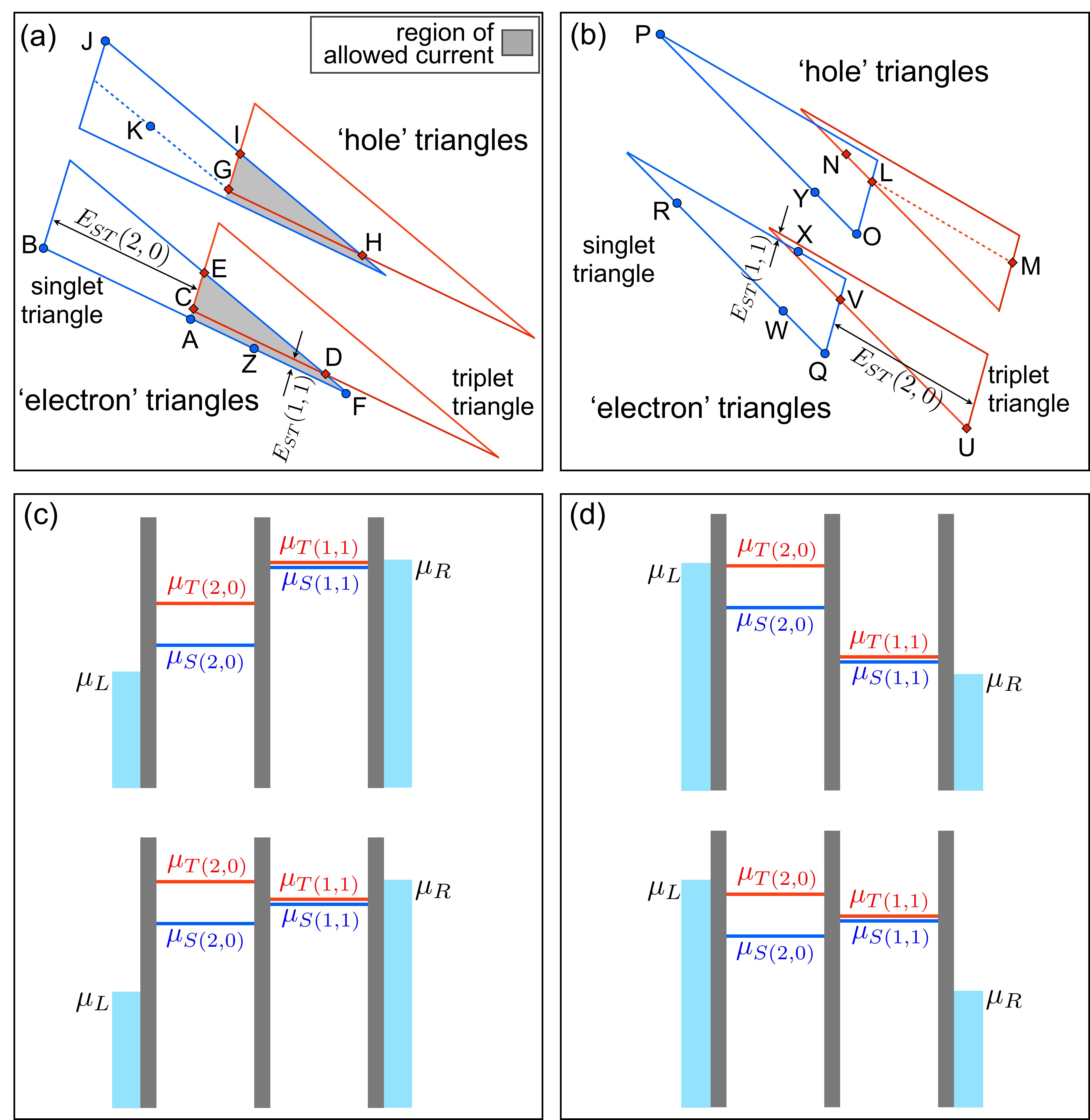}
\caption{\label{fig:triangle-schematic} 
(a) Schematic diagrams showing the singlet
  and triplet triangles for forward bias.  The gray areas show the
  region in which current is allowed in conventional spin blockade.
  In addition, current is expected along segments BA and JI, because
  on these lines the (1,1) singlet and triplet are aligned with the
  leads, and spin exchange is possible.  In the data of
  Fig.~\ref{fig:all-data}, segments BF, which has low slope, and JH,
  which has a slightly higher slope, are strongly visible and thus good candidates
  for fitting.
  (b) Schematic diagrams showing the singlet
  and triplet triangles for reverse bias.  Points U and M are outside the conventional bias triangle.
  (c), (d) Energy levels in a double quantum dot, with one fixed electron in the left-hand dot.  Panel (c) corresponds to forward bias of the the double dot, while panel (d) corresponds to reverse bias.  Electrons may transit via the singlet or triplet channels.
The upper and lower plots show the chemical potentials at two example gate voltages.  In the upper schematics of (c) and (d), electrons can transit through both the singlet and the triplet channels via an energetically downhill path.  In the lower panel of (c), spin blockade is present, as the triplet channel is energetically uphill.
The lower panel of (d) is the lifetime-enhanced transport (LET) regime, in which the singlet channel is energetically uphill, and current will be blockaded unless electrons tunnel preferentially through the triplet channel. }
\end{figure}

The data for positive source-drain bias (column one of Fig.~\ref{fig:all-data}) exhibit spin
blockade.  We refer to this bias direction as `forward bias.'  The
data in this bias direction are largely understandable using the
conventional concepts of Pauli spin blockade; 
\textcolor{black}{indeed, the consistency
of these data with classic spin blockade behavior provides strong evidence that
our assignment of the valence electron occupancies is correct.}
There are, however,
interesting resonances observable, e.g., where transport through the triplet channel of the electron
triangle overlaps with transport though the singlet channel of the hole triangle.  Also, the current in the
spin blockade regime extends by small amounts past the conventional
tips of the bias triangles. We discuss these features and others in
Sec.~\ref{sec:discussion} below.

The data for negative source-drain bias (column three of Fig.~\ref{fig:all-data}) show unusual
patterns in the current as a function of gate voltages
$V_{\mathrm{G}}$ and $V_{\mathrm{CS}}$.  We refer to this bias
direction as `reverse bias.'  Consistent with the arguments in
Sec.~\ref{sec:simple_demonstration} above and in Ref.~\onlinecite{Shaji:2008p540}, there are
regions of current extending outside the conventional bias triangle regions
--- the only regions in which transport is conventionally observed.
We have argued that current is observed outside the bias
triangles due to an effect named lifetime-enhanced transport, or LET.
The essential prerequisite for observing this effect is that long relaxation times
from an excited state, such as the (2,0) spin triplet state discussed
below, can leave open a fast, energetically downhill current path.  In
order to check this argument and understand the features shown in the
right two columns of Fig.~\ref{fig:all-data}, it is important to determine
with a fair degree of precision the sizes and positions of the bias
triangles.

The data sets in Fig.~\ref{fig:all-data}(a) and (g) were previously reported
in Ref.~\onlinecite{Shaji:2008p540}.  The
bias voltages were reported in that paper
to be \amount{+0.2}{mV} for the data in panel (a) and \amount{-0.3}{mV} for panel (g),
but these values were slightly affected by an offset in the current
preamplifier.
This offset was discovered when we
analyzed four additional data sets taken at very small bias
voltage \VSD.
To determine the size of the offset, 
we examined several data sets with small \VSD,
including Figs.~\ref{fig:simple}(a) and (b).  Cuts through the data were taken along a line connecting the two triple points (see explanation, below).  
The peak current, the full-width-at-half-maximum, and the area under
the sampled cuts were computed for each cut.  
Each of these quantities was assumed to depend linearly on \VSD~at small bias, allowing us to determine a bias offset of  0.026~mV.  We emphasize that the source-drain biases indicated in each
figure in this paper have been corrected to reflect this
offset. 

\section{Fitting}
\label{sec:fitting}

\subsection{Overview}
At infinitesimal bias voltages \VSD,
the Fermi levels of the left lead L and the right lead R are nearly equal, and current flows through the double dot only when the Fermi energies in both leads and the chemical potentials of both quantum dots are aligned, \textit{i.e.}, when $E_{LF}\simeq E_{RF}\simeq E_1\simeq E_2$.  Here, $E_{LF} (E_{RF})$ refers to the Fermi energy of the left (right) lead and $E_1 (E_2)$ refers to the chemical potential of the left (right) dot.  When these energies are equal, the charge configurations of interest are degenerate, and a `triple point' is observed in the data.\cite{VanDerWiel:2002p1382}  For the electron system, the degenerate double dot electron configurations are $(1,0)$, $(1,1)$, and $(2,0)$, while for the hole system these electron occupations are $(1,1)$, $(2,0)$, and $(2,1)$. The data in Fig.~\ref{fig:simple}(a) were acquired at a small \VSD.  As \VSD~is increased, plots of the current as a function of $E_1$ and $E_2$ (or, more conveniently, the gate voltages most directly controlling $E_1$ and $E_2$) reveal these triple points expanding into `bias-triangles' arranged regularly in the well-known pattern known as a honeycomb diagram.\cite{VanDerWiel:2002p1382}

\begin{figure}[!t]
\begin{center} \includegraphics[width=3.5in]{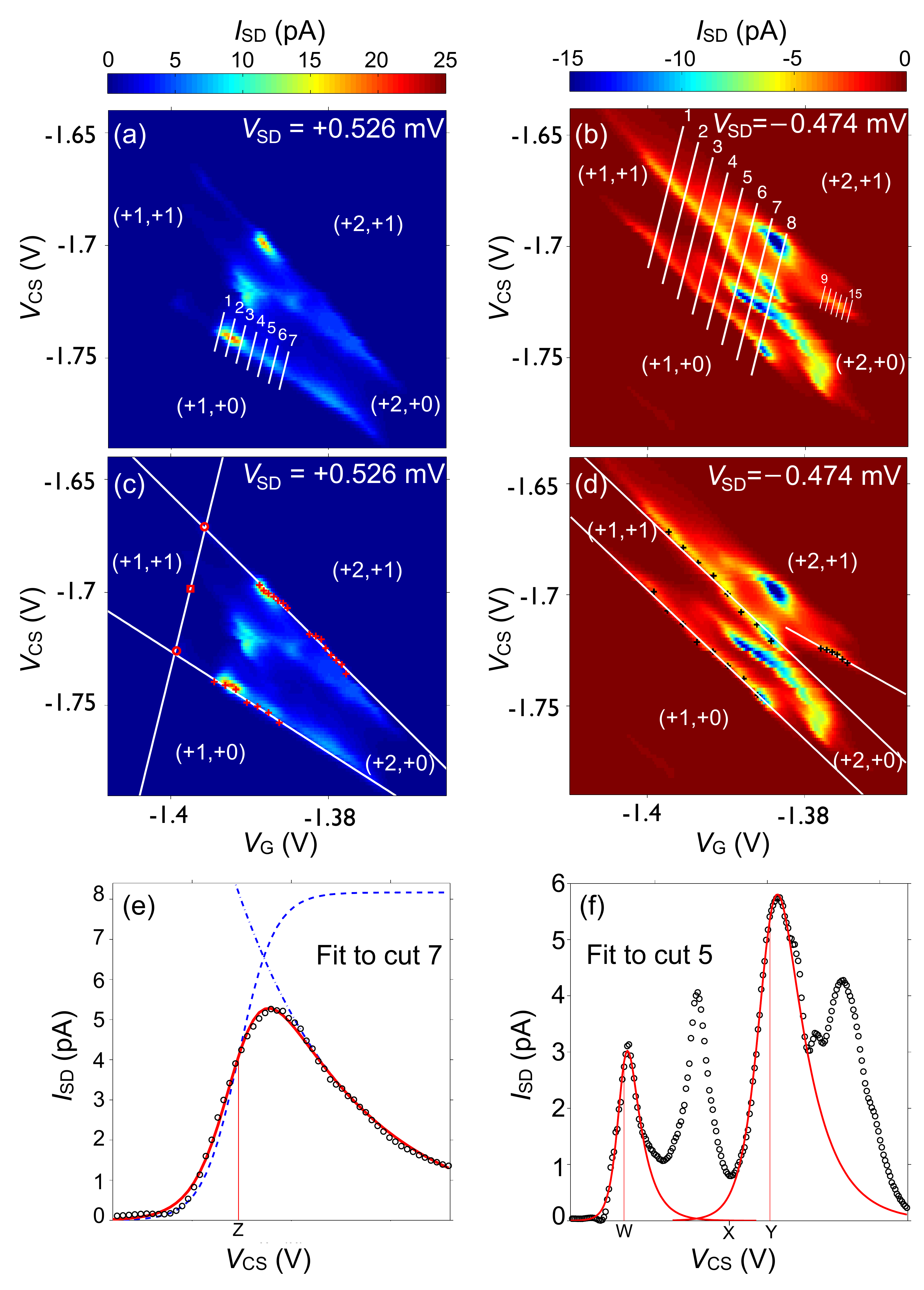} \end{center}
\caption{\label{fig:slopes-and-cuts} 
Examples showing how the triangle boundaries and
slopes are determined from the data.  
(a) Forward bias:  the positions of  the data cuts (1-7) that we use to determine the low slope. 
(b) Reverse bias:  the positions of  the data cuts that we use to determine the high slope (1-8) and the low slope (9-15). 
(c) For the same data as panel~(a), the red circles mark the points used to determine the base slope for this data set; their centroid is marked by the red square.
A white line with the mean base slope is drawn through the centroid.  
The red crosses mark points used to determine low and high triangle slopes;
the two white lines are obtained by linear fits through the crosses.
(d) For the same data as panel~(b), the black crosses mark points used to determine low and high triangle slopes;
the three white lines are obtained by linear fits through the crosses.
The separation between the triple points is labeled $d$.  
(e) Example fitting results for cut 7 in panel (a).
The dashed lines correspond to the Fermi function and the energy dependent tunnel rate, and the red line corresponds to their product.
The vertical red line indicates the triangle boundary.  
(f) Example of two independent fits to cut 5 in panel (b), and the corresponding triangle boundaries.  }
\end{figure}

The conventional bias triangle is defined by the region in gate voltage space in which the double dot ground states (in this case the singlet states) are both energetically downhill and within the source-drain bias window.  We refer to these as the `singlet triangles.'  There are analogous `triplet triangles' defined by similar conditions for the excited triplet states.  
Separate singlet and triplet triangles can be defined for both the electron and hole systems.
For electrons in forward bias, transport occurs through the sequence (1,0) $\rightarrow$ (1,1) $\rightarrow$ (2,0), while for the hole regime, in forward bias, transport occurs through the
electron sequence (2,1) $\rightarrow$ (1,1) $\rightarrow$ (2,0).  Figures~\ref{fig:triangle-schematic}(a) and (b) show schematic diagrams of these four distinct triangles for the cases of forward bias in panel (a) and reverse bias in panel (b).  Note that the lengths BA and AC represent the singlet-triplet energy splittings in the (2,0) and (1,1) states, respectively.
\textcolor{black}{The magnitude of the measured
current varies significantly across each triangle because of
energy-dependent tunneling;\cite{Petta:2005p161301,Ono:2005p1313,Johnson:2005p925,Johnson:2005p483,Maclean:2007p1499,Amasha:2008p1500} the tunneling
rate from the lead decreases as the tunnel barrier increases, which occurs as the
energies of the relevant levels in the dots are lowered below the Fermi level.
The effects of energy-dependent tunneling are more marked in Si/SiGe dots
than in GaAs dots, because electrons in silicon have 
larger effective mass,\cite{DaviesBook} and these effects are discussed in more
detail below, in Sec.~\ref{sec:theory}. }

Points Q, R, X, W, and V in Fig.~\ref{fig:triangle-schematic}(b) lie within the conventional bias triangle.  Point U lies outside this triangle, but within the triplet triangle.  Fig.~\ref{fig:triangle-schematic}(d) shows example energy level arrangements for points within both the singlet and triplet triangles (upper cartoon), and points such as U, that lie outside the singlet triangle but within the triplet triangle (lower cartoon).   In the conventional picture, without LET, the regions of strong current flow are the blue singlet triangles, while the regions of the red triplet triangles that do
not overlap the singlet triangles are blockaded.  A major point of this paper is to demonstrate that the current visible in the lower right hand corners of the panels in column three of Fig.~\ref{fig:all-data} arises because significant current is flowing through the triplet states outside the conventional bias triangle.  Because of its shape, we refer to this feature in the data as the triplet `tail.'  LET lifts the blockade condition, giving rise to the triplet tail when tunnel rates and triplet-singlet relaxation times are appropriate \textcolor{black}{---
specifically, the triplet must load enough faster than the singlet that a transport current
is measurable, even though 
tunneling through the singlet channel is very slow}.\cite{Shaji:2008p540}

In order to understand how electrons move from one lead to the other through the
double dot system, it is necessary to know the chemical
potentials of each quantum dot for particular occupancy states. Of
particular significance are the chemical potentials associated
with points labeled with uppercase letters in Fig.~\ref{fig:triangle-schematic}(a) and (b).  The corresponding energy level diagrams are all presented in the appendix.  To understand the data in columns one and three of
Fig.~\ref{fig:all-data}, simultaneously and self-consistently, we aim to determine the size and shape of the bias
triangles, and to position these triangles as accurately as possible
on the data.  The end result of this analysis is shown in the second
and fourth columns of Fig.~\ref{fig:all-data}.

Our procedure is as follows: first, using data for all biases and from
both the electron and hole systems, we obtain the three slopes
that define the edges of the bias triangles \textcolor{black}{(see Fig.~1(f))}; these are (i) the slope of the base of the
triangle (base slope), which characterizes the direction in gate voltage space in which
the chemical potentials of the two dots are held constant relative to each other,
(ii) the low slope of one of the long edges (low slope), which characterizes the direction in gate voltage space in which
the chemical potential of the right dot is held constant with respect
to the Fermi level of the right lead,
and (iii) the slightly higher slope of the other long edge (high
slope), which characterizes the direction in gate voltage space in which
the chemical potential of the left dot is held constant with respect
to the Fermi level of the left lead.  Second, we determine the separation between the electron and
the hole triangles, as well as the scaling relation between the applied bias voltage
and the size of the triangles.  Finally, we position the triangles on
the data sets in Fig.~\ref{fig:all-data}.

\subsection{\label{sec:base}Determination of the base slope}

The base slope joins the points B and J in Fig.~\ref{fig:triangle-schematic}(a) and, similarly, the points Q and O in Fig.~\ref{fig:triangle-schematic}(b). 
Points B and J are identified as small, solitary peaks in the left-hand region of the forward bias experimental data [column 1 of
Fig.~\ref{fig:all-data} and expanded in Fig.~\ref{spinex}(a, b)]. Note that the data were interpolated using the cubic spline procedure, \footnote{In fitting the
  data, two processing steps were performed: (i) cubic spline interpolation
  of five points for every single data point was used, and (ii) the
  average background current was determined and subtracted from each
  data set.} to more accurately identify the center of the points, while keeping the functional form
of the data unchanged. The base slopes were obtained for
each pair of points (B,J), in all four forward bias data sets, and the average value was calculated, with results summarized in Table~\ref{tab:bslopes}.
The base line for each forward bias data set is obtained by determining the centroid position of a given (B,J) pair, and then drawing a line with the average base slope through the centroid.
An example of this procedure is shown in Fig.~\ref{fig:slopes-and-cuts}(c), with the centroid indicated as a red square and the points B and J indicated by the red circles.

\begin{table}[htdp]
  \caption{\label{tab:bslopes}Base slopes obtained from the spin
    blockade data of Fig.~\ref{fig:all-data}.}
\begin{ruledtabular}
\begin{center}
\begin{tabular}{ c c}
Bias (mV)&Base slope\\[0.5ex]
\hline \\[-2.4ex]
0.226 & 15.3 \\
0.326 & 11.2 \\
0.526 & 15.5 \\
0.626 & 14.8 \\[0.2ex]
\hline \\[-2ex]
\multicolumn{2}{c}{$\text{Base slope mean} = 14.2$}\\
\multicolumn{2}{c}{Standard deviation = 2.02}\\
\end{tabular}
\end{center}
\end{ruledtabular}
\end{table}

\subsection{Determination of the low slope}
Both the forward and reverse bias data possess features that are
useful for determining the low slope of the bias triangles. 
We first consider the forward bias data along the line segment AF shown in Fig.~\ref{fig:triangle-schematic}(a).
Along this line, the chemical potential of the left dot is variable, while the chemical potential of the right dot is constant.
The current flow is nearly constant along AF, except for the resonant peak, which we will discuss in Section~\ref{sec:theory}.
Thus, the current depends only weakly on the chemical potential of the left dot.
We then take data cuts parallel to the base slope, crossing the line segment AF, as shown in Fig.~\ref{fig:slopes-and-cuts}(a).  
Along a given cut, the current does not flow uniformly and it does not fill up the whole bias triangle.
The strong suppression of the current above AF is a signature of energy dependent tunneling.  
We conclude that the current depends most strongly on the chemical potential of the right dot, and that the right-hand tunnel barrier forms the transport bottleneck.

In Section~\ref{sec:data_energy} below, we provide detailed models for energy dependent tunneling.
However, in order to delineate the edges of the bias triangle here, we simply point out that the dominant 
contribution to the current along the data cuts in Figs.~\ref{fig:slopes-and-cuts}(a) and (b) can be expressed as follows:
\begin{equation}
I/e=f_{R}\Gamma_{R} .
  \label{eq:1}
\end{equation}
Here, 
\begin{equation}
f_{R}=\left[ e^{(E-E_{RF})/k_BT}+1 \right]^{-1}
  \label{eq:2}
\end{equation}
is the Fermi function for the right lead, where $E$ is the chemical potential of the right dot, and $E_{RF}$ is the Fermi level of the lead.
The Fermi function defines the edge of the forward bias triangle along the line AF, defined by the condition $E=E_{RF}$.  
The second function appearing in Eq.~(\ref{eq:1}) is the effective tunneling rate $\Gamma_R$, from the right lead to the right
dot. To capture the effect of energy dependent tunneling, we will apply approximations similar to those used in
Refs.~\onlinecite{Maclean:2007p1499} and \onlinecite{Amasha:2008p1500}, and discussed in greater detail in
Sec.~\ref{sec:data_energy}, leading to the prescription
\begin{equation}
\Gamma_{R}=\Gamma _{R0} e^{(E-E_{RF})/E_{R0}},
  \label{eq:3}
\end{equation}
where $\Gamma_{R0}$ is proportional to the attempt rate, and $E_{R0}$ describes the scale for the energy dependent tunneling.
This exponentially decaying function suppresses the current flow above line segment AF.

Equation~(\ref{eq:1}) can be used to fit the data along the cuts shown in Fig.~\ref{fig:slopes-and-cuts}(a) by 
assuming a linear relation between the chemical potential of the right dot and the control voltages $V_\textrm{CS}$ and $V_\textrm{G}$.
The proportionality constants, the so-called `lever-arms', are determined as part of the fit.  
A typical result of the fitting procedure is shown in Fig.~\ref{fig:slopes-and-cuts}(e). The
vertical line on the plot represents the boundary of the bias triangle, corresponding to the condition $E=E_{RF}$. 
The boundary positions for each of the cuts in Fig.~\ref{fig:slopes-and-cuts}(a) were obtained in the same way, giving the points marked as red crosses in Fig.~\ref{fig:slopes-and-cuts}(c).  The edge 
of the bias triangle was then determined by fitting a straight line through these points, with the result shown in Fig.~\ref{fig:slopes-and-cuts}(c)

The low slope can also be investigated in the hole system in the reverse bias regime, along the line segment LM shown in Fig.~\ref{fig:triangle-schematic}(b).  Data cuts are again taken parallel to the base slope, as shown by the short lines on the right-hand-side of 
Fig.~\ref{fig:slopes-and-cuts}(b). 
A function similar to Eq.~(\ref{eq:1}) was used to fit the data, to obtain the triangle boundary along each cut.   The results are shown as
black crosses superimposed on the right-hand side of Fig.~\ref{fig:slopes-and-cuts}(d).  Again, we fit a straight line through these boundary points, obtaining the result shown in panel (d).  The low slopes were obtained in this way for 8 different data sets, as summarized in Table~\ref{tab:lslopes}.

\begin{table}[htdp]
  \caption{\label{tab:lslopes}Low slopes obtained from both forward and
    reverse bias data.}
\begin{ruledtabular}
\begin{center}
\begin{tabular}{ c c | c c}
 \VSD~(mV)&Low slope& \VSD~(mV)&Low slope\\[0.5ex]
  \hline \\[-2.4ex]
  0.226 & -2.25 & -0.174 &-1.94\\
  0.326 & -2.17 & -0.274 & -1.93\\
  0.526 & -2.24 & -0.474 & -1.97\\
  0.626 & -2.24 & -0.574 & -1.95\\[0.2ex]
  \hline \\[-2ex]
  \multicolumn{4}{c}{Low slope mean = -2.09}\\
  \multicolumn{4}{c}{Standard deviation= 0.15}\\
\end{tabular}
\end{center}
\end{ruledtabular}
\end{table}

\subsection{Determination of the high slope and the triple point spacing}
The high slope can also be determined from fits to both forward and reverse bias data.  
In the case of forward bias, we take data cuts though the hole data, along lines parallel to the base slope.  
Along the line IH, indicated in Fig.~\ref{fig:triangle-schematic}(a), the chemical potential of the left dot is equal to the Fermi level of the left lead.
For the hole triangle, the energy dependent tunneling is rather weak.
The prominent features in the data along line segment IH are mainly attributed to the Fermi function for the left lead.
We therefore adopt the following fitting form for data near line segment IH:
\begin{equation}
I/e=(1-f_L)\Gamma_L ,
\end{equation}
where the Fermi function $f_L$ and the energy dependent tunnel rate $\Gamma_L$ are defined analogously to Eqs.~(\ref{eq:2}) and (\ref{eq:3}). 
The data cuts are fit as described above, for source-drain biases $V_\mathrm{SD}=0.226$, 0.526, and 0.626~mV.  
The 0.326~mV data set in Fig.~\ref{fig:all-data}(e) exhibits a discontinuity along IH arising from a charging event.
For that data set alone, the fitting procedure is performed in the vicinity of line segment JI rather than IH.
The fitting results for the triangle boundaries are shown as red crosses in Fig.~\ref{fig:slopes-and-cuts}(c), for each data cut.  
The high slope is obtained from a linear fit through the boundary points, as given by the white line.  

In the reverse bias regime, we also take data cuts parallel to the base slope.  
In this case, the cuts extend across the entire electron-hole system, as shown in Fig.~\ref{fig:slopes-and-cuts}(b).
We use form
\begin{equation}
I/e=f_L\Gamma_L 
\end{equation}
to fit the data.
In this case, however, independent fits are made to both the electron and the hole peaks, giving typical results as shown in Fig.~\ref{fig:slopes-and-cuts}(f).
Here, the vertical lines represent the inferred locations of both triangle boundaries.
The resulting boundary locations are shown as black crosses in Fig.~\ref{fig:slopes-and-cuts}(d).
By fitting straight lines, we obtain results for the high slope, as summarized in Table~\ref{tab:hslopes}.

The fits to the reverse bias data provide a direct method for determining the separation between the electron and hole triangles along the direction parallel to the base slope. 
This separation is the triple point spacing, and it is indicated by the distance $d$ in Fig.~\ref{fig:slopes-and-cuts}(d). 
It is also indicated, schematically, by the distance between points Q and O in Fig.~\ref{fig:triangle-schematic}(b). 
The triple point spacing is the same for all biases, and we therefore determine its value by averaging the individual extracted values for $d$.

\begin{table}[htdp]
  \caption{\label{tab:hslopes}High slopes obtained from both forward and
    reverse bias data.}
\begin{ruledtabular}
\begin{center}
\begin{tabular}{ c c | c c}
 \VSD~(mV)&High slope&\VSD~(mV)& High slope\\[0.5ex]
  \hline \\[-2.4ex]
  0.226 & -3.32 & -0.174 &-3.86\\
  0.326 & -3.75 & -0.274 & -4.06\\
  0.526 & -3.47 & -0.474 & -3.37\\
  0.626 & -3.59 & -0.574 & -3.71\\[0.2ex]
 \hline \\[-2ex]
  \multicolumn{4}{c}{High slope mean = -3.64}\\
  \multicolumn{4}{c}{Standard deviation= 0.25}\\
\end{tabular}
\end{center}
\end{ruledtabular}
\end{table}

\section{Positioning and scaling the triangles}
\label{sec:positioning}
It is now possible to draw the singlet and triplet bias triangles.  The shape of each triangle is known precisely in terms of the high, low, and base slopes.  In this section, we explain how the sizes of the triangles are determined and how they are positioned on the data.

The size of the triangles is proportional to the source-drain bias.  To determine the scaling, we focus on the forward bias data sets.  By using the triple point spacing $d$, in combination with line fits of the type shown in Fig.~\ref{fig:slopes-and-cuts}, the triangles for the forward bias data are completely determined.  We extract a gate voltage-to-energy proportionality constant for each forward bias data set for both $V_\mathrm{CS}$ and $V_\mathrm{G}$, and we calculate the mean values of each.  We use these constants to set the size of the triangles for the reverse bias data. The sizes of the triangles drawn in column 4 of Fig.~\ref{fig:all-data} are all obtained using these mean voltage-to-energy calibrations.

We now position the triangles in the forward bias regime.  Since line segment CD cannot be easily distinguished from AF in the forward bias data, the triplet triangles are initially positioned by assuming that AF and CD overlap.  That is, we assume zero $(1,1)$ singlet-triplet energy splitting. The actual $(1,1)$ singlet-triplet splitting is determined later.  We then determine the base position of the triplet triangle by performing a Lorentzian fit to a data cut along line segment BF, placing the triangle corner at the peak of the Lorentzian.

To position the singlet triangles on the reverse bias data plots, we only need to determine a single point, which we take to be point Q in Fig.~\ref{fig:triangle-schematic}(b).  This point is determined for each data set by fitting a Lorentzian to a data cut along line segment RQ.  Similarly, the triplet triangles are positioned by performing a Lorentzian fit to point U along the line segment VU.  This method of independently positioning the singlet and triplet triangles provides a means of estimating the (1,1) singlet-triplet splitting, which is observed as a slight shift in the edges of the triangles.
Note that, due to the lack of a straight line of current for the two higher bias data sets, we use the coordinates of the strongest current peak in that vicinity to locate point U. This should lead to an overestimate of the (1,1) singlet-triplet energy splitting in those two cases.

The (2,0) singlet-triplet splitting is given by the distance between the bases of the singlet and triplet triangles.  These voltages are converted to energies using our lever-arm calibrations.  We then take a mean value from all eight data sets, obtaining the result $E_{ST}(2,0)=0.173$~meV, as reported in Table~\ref{tab:est}.  This value is interesting, because it corresponds to the lowest excited state that is not a spin excitation.  In general, this degree of freedom will involve an orbital parameter, and specifically, it can correspond to the valley degree of freedom.  As such, the mean value of $E_{ST}(2,0)$ provides a lower bound on the valley splitting \cite{Boykin:2004p115,Boykin:2005p1461,Friesen:2006p202106,Friesen:2007p115318,Kharche:2007p092109,Goswami:2007p41} in dot~1.  

We can compute the singlet-triplet splittings $E_{ST}(1,1)$ in a similar manner.  Such estimates can only be obtained in the LET data sets, since the forward bias data sets do not provide a good signature of the splitting.  We find that the standard deviation of $E_{ST}(1,1)$ is almost as large as its mean value.  Such a large uncertainty in $E_{ST}(1,1)$ is not surprising, because its value is similar to the estimated electron temperature during these measurements, which is 145~mK.  The resulting estimates for $E_{ST}(1,1)$ is reported in Table~\ref{tab:est}.

\begin{table}[htdp]
\caption{\label{tab:est}Singlet-triplet energy splitting.}
\begin{ruledtabular}
\begin{center}
\begin{tabular}{c c c}
\VSD~(mV)&$E_{ST}(1,1)$ (meV)&$E_{ST}(2,0)$ (meV)\\[0.5ex]
\hline \\[-2.4ex]
-0.174 &  0.0078 &0.170 \\
-0.274 &  0.0044 & 0.174 \\
-0.474 &  0.0306 &0.148 \\
-0.574 & 0.0345 &0.193 \\
0.226 &   & 0.200\\
0.326 & & 0.178 \\
0.526 &  & 0.155\\
0.626 & & 0.162\\
\hline \\[-2.4ex]
Mean  &  0.019 &  0.173 \\
Standard deviation & 0.015 &  0.018\\[-0.5ex]
\end{tabular}
\end{center}
\end{ruledtabular}
\end{table}

\section{Theoretical Model for Energy-Dependent Tunneling Effects}
\label{sec:theory}

In this section, we investigate processes related to LET in the reverse bias regime, and we analyze energy-dependent tunneling and its impact on the transport.
We focus specifically on the lower two triangles of Fig.~\ref{fig:triangle-schematic}(b).
In this case, one valence electron is always present in the left dot, while a second valence electron transits the double dot from the left to the right, in the charging sequence $(1,0)\rightarrow (2,0)\rightarrow (1,1)\rightarrow (1,0)$.  Note that a downhill energy path between the left and right dots corresponds to a positive value of the detuning parameter,  $\varepsilon \equiv (E_1-E_2)>0$, where $E_1$ and $E_2$ are the chemical potentials of the left dot (dot 1) and the right dot (dot 2), respectively.

\subsection{Qualitative Discussion}

In Ref.~\onlinecite{Shaji:2008p540}, a sequential tunneling model was used to analyze the reverse bias transport currents.  In the sequential tunneling approximation, the current through a particular transport channel can be expressed as $(I/e)^{-1}=\Gamma_{L}^{-1}+\Gamma_{12}^{-1}+\Gamma_{R}^{-1}$, where the $L$ index refers to the $\mathrm{L} \rightarrow 1$ tunnel process, the R index refers to the $2\rightarrow R$ process, and the $12$ index refers to tunneling between dots 1 and 2.  
As in previous sections, $L$ and $R$ refer to the left and right leads.  In Ref.~\onlinecite{Shaji:2008p540}, two transport channels were studied:  the singlet channel (S) and the triplet channel (T).

The sequential tunneling model provides a great deal of information.  For example, it 
explains why the lower portions of the data in Fig.~\ref{fig:all-data}(c) and (g) take the distinctive form of two parallel lines, rather than a triangle: the two lines correspond to distinct transport processes through the singlet and triplet channels.  For either channel, the current is effectively determined by the bottleneck process, which turns out to be $\Gamma_{L}$ or $\Gamma_{12}$.  
Since the function $\Gamma_{L}$ depends sensitively on the chemical potential of dot 1, due to energy dependent tunneling,
we observe that $I$ is exponentially suppressed when $E_1< E_{LF}$,
reducing the bias `triangle' to a narrow line.  
Thus, the lower current feature in Fig.~\ref{fig:triangle-schematic}(b) actually consists of two overlapping triangles (a singlet triangle and a triplet triangle), each of which is reduced to a narrow line due to energy dependent tunneling.

Despite its success, the sequential tunneling model is over-simplified and cannot explain certain crucial features of the transport current.  For example, the strong enhancements of the transport current at the points marked Q and U in Fig.~\ref{fig:triangle-schematic}(b) are resonances arising from the coherent delocalization of electrons in dots 1 and 2.  Such effects cannot be explained by an incoherent tunneling model.  The master equation approach of Nazarov and Stoof does incorporate resonant effects.\cite{Nazarov:1993p57,Stoof:1996p1050}
However, it does not account for the inelastic, sequential tunneling processes that dominate the transport throughout most of the current map.  
\textcolor{black}{We use a
master equation technique that incorporates both resonant and inelastic tunneling effects
to address this situation.}

\subsection{Quantitative Analysis}
\textcolor{black}{The theoretical model that we use for the quantitative analysis
is presented in detail in Appendix~\ref{lindblad}.
This model treats the S and T transport channels independently,
and treats the coupling to the environment within the Lindblad 
formalism.\cite{PreskillNotes,NielsenBook}
The analysis yields an expression for the
current $I$ through a single channel in the two-electron dot:
\begin{widetext}
\begin{equation}
I= e\Gamma_L\Gamma_R \frac{ \begin{array}{l}
\left\{ 4t^2(f_L-f_R)[\Gamma_L(1-f_L)+\Gamma_R(1-f_R)+\Gamma_i] 
\right. \\ \left. \hspace{1in}
+ \Gamma_i\left( 4(\varepsilon/\hbar)^2+[\Gamma_L(1-f_L)+\Gamma_R(1-f_R)+\Gamma_i] ^2 \right)
[f_L(1-f_R)\theta -f_R(1-f_L)\bar\theta ] \right\} \end{array} }
{\begin{array}{l}
\left\{ \left[ 4(\varepsilon/\hbar)^2+[\Gamma_L(1-f_L)+\Gamma_R(1-f_R)+\Gamma_i]^2 \right]
[\Gamma_L\Gamma_R(1-f_Lf_R)+\Gamma_i\Gamma_R(\theta+f_R\bar\theta)
+\Gamma_i\Gamma_L(\bar\theta +f_L\theta)] 
\right. \\  \left. \hspace{1in}
+4t^2 [\Gamma_L(1-f_L)+\Gamma_R(1-f_R)+\Gamma_i]
[\Gamma_L(1+f_L)+\Gamma_R(1+f_R)]  \right\} 
\end{array} } . \label{eq:full}
\end{equation}
\end{widetext}
Here, $t$ is the (elastic) tunnel coupling between the two dots,
$\varepsilon$ is the energy difference between the (2,0) charge configuration
and the (1,1) charge configuration,
$f_L$ and $f_R$ are the Fermi functions for the two leads $L$ and $R$ (they of
course depend on energy, but this dependence is suppressed in the notation
for brevity),
$\Gamma_L$  is the tunnel coupling between the left lead $L$ and dot~1,
$\Gamma_R$ is the tunnel coupling between the right lead $R$ and dot~2,
$\Gamma_i$ is the inelastic interdot coupling,
and $\theta$ and $\bar\theta$, which account for the fact that phonon emission is much more
likely than phonon absorption at low temperatures,
are taken here to be Heaviside step functions:
$\theta =\Theta(\varepsilon)$ and $\bar\theta =\Theta(-\varepsilon)$, 
with $\Theta(\varepsilon)=0$ when $\varepsilon<0$, 
$\Theta(\varepsilon)=1/2$  when 
$\varepsilon=0$, and $\Theta(\varepsilon)=1$ when $\varepsilon>0$. 
The inelastic interdot tunnel coupling $\Gamma_i$ is a weak, even function of $\varepsilon$.  Below, we show that our data are consistent with $\Gamma_i=(\text{constant})$ throughout most of the bias triangle.
When the singlet and triplet channels are fully decoupled, as we assume here, the total current is expressed as a sum of terms like Eq.~(\ref{eq:full}), one for each channel.  
For the case of triplet states that are triply degenerate, the total triplet current is therefore the sum of three terms, one for each state. 
}

\begin{figure}[!t] 
  \centering
  \includegraphics[width=2.2in,keepaspectratio]{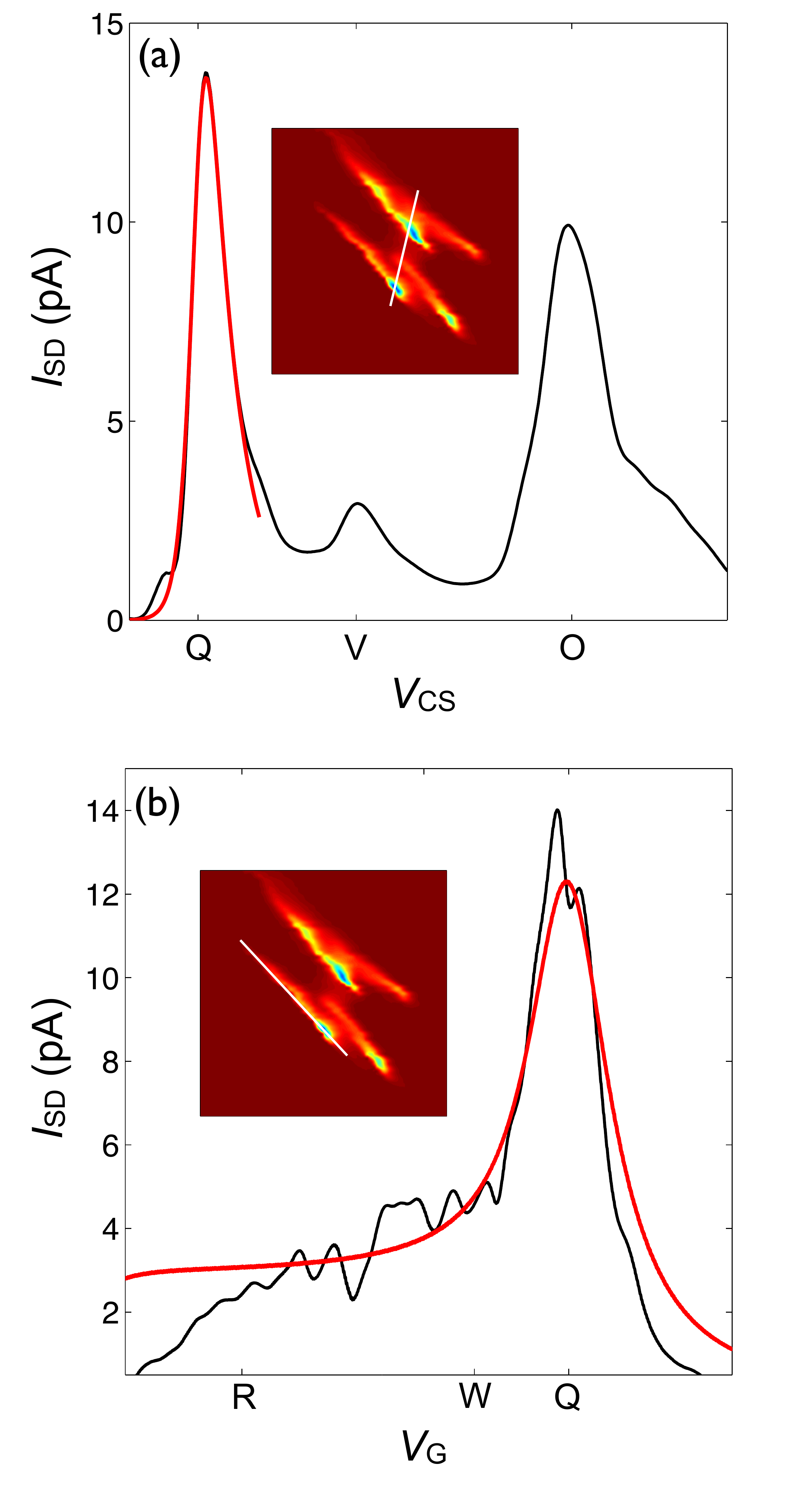}
  \caption{
Fits to data cuts.
(a) Black curve:  a cut through the data obtained along the line segment QVO, as indicated in the inset. 
Red curve:  a fit to the 1D data cut, using the theoretical formula in Eq.~(\ref{eq:res2}).
(b) Black curve: a cut through the data along the line segment RWQ, as indicated in the inset.
Red curve:  a cut through the theoretical fit to the 2D data set, evaluated along the same line.  The data were fit using the full theoretical model of Eq.~(\ref{eq:full}).}
  \label{fig:fits-to-cuts}
\end{figure}

\section{Analysis of data yielding information about energy-dependent tunneling}
\label{sec:data_energy}
In this section, we first perform a fitting analysis using Eq.~(\ref{eq:full}) to obtain estimates for the various tunneling parameters, including the energy dependent tunneling.  We then go on to discuss the prominent features in the current map.  We finish up by checking the self-consistency of our LET assumption of decoupling between the single and triplet channels, and we discuss the implications for preferential loading of the excited states.

\subsection{Fitting analysis of the tunnel parameters}
We begin with an investigation of the various tunnel rates in the LET regime.
The transport data of Fig.~\ref{fig:all-data}(g) will be analyzed along particular cuts.  We first consider the cut QVO, which is along the base of the singlet triangle, as shown in Fig.~\ref{fig:fits-to-cuts}(a). 
We also consider the data cut RWQ, which is along the high slope of the singlet triangle, as shown in Fig.~\ref{fig:fits-to-cuts}(b).  Along the latter cut, the data exhibit two prominent features:  a Lorentzian peak, which is characteristic of resonant tunneling, and a relatively flat region to the left of the peak, which is characteristic of inelastic tunneling.\cite{Fujisawa:1998p932}   

We first analyze the singlet inelastic transport current, which dominates the current flow over most of the singlet bias triangle, except near the line QV.  As we shall see, the tunnel coupling $t$ has a characteristic magnitude of $\mu$eV, while the length of the bias triangle is on the order of hundreds of $\mu$eV, in energy units.  Thus, away from line segment QV, the condition $\varepsilon \gg \hbar t$ is true almost everywhere.
Equation~(\ref{eq:full}) then reduces to the expected form for sequential tunneling
\begin{equation}
I /e\simeq f_L(1-f_R)\left[ \Gamma_L^{-1}+\Gamma_R^{-1}+\Gamma_i^{-1} \right]^{-1} ,
\label{eq:seq}
\end{equation}
where the tunneling between dots 1 and 2 is strictly inelastic.  

To make further progress,
it is useful to introduce a specific model for the tunneling rates between the dots and the leads.  For simplicity, we consider square tunnel barriers, for which the leading order energy dependence of the tunnel rate is exponential and is given by \cite{DaviesBook}
\begin{equation}
\Gamma (E)=\Gamma_0 e^{-2W\sqrt{2m^*(U-E)/\hbar^2}} . \label{eq:square}
\end{equation}
Here, $W$ is the barrier width, $U$ is its height, and $E$ is the energy of the tunneling electron.
Since our transport data do not exhibit enough structure to independently determine the parameters characterizing the tunnel barriers, we consider an alternative tunneling function by linearizing the argument of the exponential in Eq.~(\ref{eq:square}) about one of the lead Fermi levels (see Refs.~\onlinecite{Maclean:2007p1499} and \onlinecite{Amasha:2008p1500}).  
For the tunneling function between the left lead and dot~1, we perform our expansion around the Fermi energy of the left lead, obtaining
\begin{equation}
\Gamma_L(E_1)\simeq \Gamma_{L0} e^{(E_1-E_{LF})/E_{L0}} , \label{eq:linear}
\end{equation}
with the characteristic energy defined as $E_{L0}=[(U_{L}-E_{LF})\hbar^2/2m^*W_{L}^2]^{1/2}$.  An analogous linearization can also be performed for the right lead.  

We first consider the singlet triangle.  We can make a rough comparison of the magnitudes of the different tunnel rates in Eq.~(\ref{eq:seq}), based on general observations of  the data in Fig.~\ref{fig:fits-to-cuts}.  We first consider the flat region near point W in panel (b).
To the left of this region, the bias triangle closes, due to the action of the Fermi functions.  To the right, we observe the resonant peak at point Q.  Along the line segment RQ, the chemical potential of the left dot is constant, so $\Gamma_L$ must be constant, but $\Gamma_R$ need not be.  Since the data are almost flat, $\Gamma_R$ must not determine the shape of the current flow.
For data cuts parallel to line segment QVO, the detuning parameter $\varepsilon$ is a constant, so $\Gamma_i$ must be almost constant.  However, the current has a strong energy dependence, which cannot be due to $\Gamma_i$.
Together, these facts suggest that the energy dependence of $\Gamma_L$ controls the shape of the current in the inelastic tunneling regime, although not necessarily its magnitude.  We conclude that the functions $\Gamma_R$ and $\Gamma_i$ must either be much larger than $\Gamma_L$ or constants in the inelastic tunneling regime.

We now perform a more quantitative analysis by considering the line QV, defined by the resonant condition $\varepsilon= 0$.
Because of the resonance, terms involving $t$ must be dominant in Eq.~(\ref{eq:full}).
Away from the long edges of the triangle, Eq.~(\ref{eq:full}) then reduces to 
\begin{equation}
I/e \simeq (\Gamma_L^{-1}+2\Gamma_R^{-1})^{-1} . \label{eq:res1}
\end{equation}
As expected, we find that the inelastic tunneling contribution, $\Gamma_i$, is irrelevant in the resonant regime.
This fact makes it possible to independently determine the parameters $\Gamma_R$ and $\Gamma_i$.  By comparing Eqs.~(\ref{eq:seq}) and (\ref{eq:res1}) and noting that the current in Fig.~\ref{fig:fits-to-cuts}(b) is much larger at point Q than point W, we conclude that $\Gamma_i\ll \Gamma_R$ in the inelastic tunneling regime.
This fact is not affected by the resonance condition.  Since $\Gamma_L$ corresponds to the bottleneck process in the resonant tunneling regime, we find that $\Gamma_L,\Gamma_i\ll \Gamma_R$.  Equation~(\ref{eq:full}) then reduces to
\begin{equation}
I/e\simeq \Gamma_L(f_L-f_R) . \label{eq:res2}
\end{equation}

We can fit Eq.~(\ref{eq:res2}) to the data cut along QV, as shown in Fig.~\ref{fig:fits-to-cuts}(a).  This gives a direct estimate for the temperature and the energy dependent tunneling parameters in the linearized function $\Gamma_L$.  We can obtain the remaining singlet tunneling parameters by performing a 2D fit of the data to the full expression in Eq.~(\ref{eq:full}).  This provides estimates for the parameters $\Gamma_i$, $\Gamma_R$ and $t$, with results shown in Table~\ref{tab:table1}.  We note that since $\Gamma_R$ has been proven to be irrelevant in our LET data, it was not possible to discern any energy dependence in this parameter.  Thus, we have treated $\Gamma_R$ as a constant in our analysis.  In Fig.~\ref{fig:fits-to-cuts}(b), we show one result from our 2D fitting procedure, as evaluated along the line segment RWQ .

\begin{table}
\caption{\label{tab:table1}
Fitting parameters and singlet-triplet energy splittings for the Si/SiGe double quantum dot transport model presented in Eq.~(\ref{eq:full}) for the data in Fig.~\ref{fig:all-data}(g).  The energy dependent tunneling parameters and the singlet-triplet energy splittings are described in the text. Standard deviations are given in square brackets.}
\begin{ruledtabular}
\begin{tabular}{lc}
& Energy ($\mu$eV) \\[0.5ex]
\hline \\[-2.2ex]
$h\Gamma_{L0S}$ & 0.62 [0.01] \\
$E_{L0S} $ & 40 [2] \\
$h\Gamma_{iS}$ & 0.125 [0.003] \\
$\hbar t_S$ & 3.2 [1.2] \\
$h\Gamma_{RS}$ & 38 [28] \\[0.5ex]
\hline \\[-2.2ex]
$h\Gamma_{L0T}$ & 0.48 [0.01] \\
$E_{L0T} $ & 34 [3] \\
$h\Gamma_{iT}$ & 0.183 [0.003] \\
$\hbar t_T$ & 2.0 [0.1] \\
$h\Gamma_{RT}$ & 55 [8] \\[0.5ex]
\hline \\[-2.ex]
Temperature & 145 [7] mK\\[0.5ex]
\hline \\[-2.2ex]
$E_{ST}$, (2,0) state & 174 [38]\\
$E_{ST}$, (1,1) state & 4 [1]\\ 
\end{tabular}
\end{ruledtabular}
\end{table}

\begin{figure}[t] 
  \centering
  \includegraphics[width=2.5in,keepaspectratio]{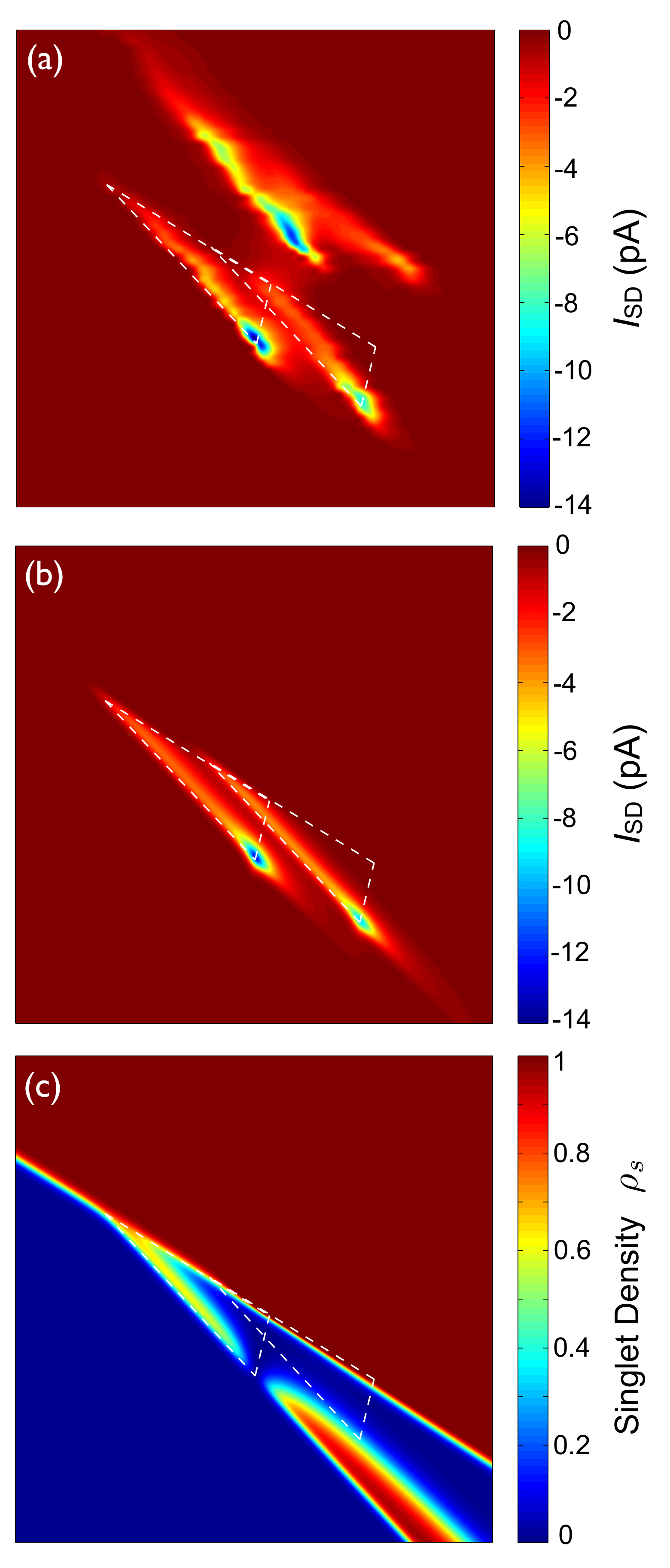}
  \caption{
 Comparison of the experimental data with calculations based on the fitting parameters of Table~\ref{tab:table1}, obtained for the same range of gate voltages.
(a)  Current transport data, identical to Fig.~\ref{fig:all-data}g, with an overlay of the edges of the singlet and triplet lower bias triangles.
(b)  Theoretical reconstruction of the lower bias triangles, based on Eq.~(\ref{eq:full}).
(c)  The computed singlet occupation density, $\rho_S=\rho_{1S}+\rho_{2S}$, shows that the singlet occuption falls off in the vicinity of the triplet triangle, as required for the observation of LET.}
  \label{fig:reconstruct}
\end{figure}

\subsection{Prominent features in the data}
It is instructive to consider limiting cases of Eq.~(\ref{eq:full}) that are relevant for our LET data, in order to gain a better physical understanding.  We specifically consider the bright line of current adjacent to line segment RWQ.  As apparent from Fig.~\ref{fig:slopes-and-cuts}(e), the Fermi function for the left lead is nearly saturated along this line, so that $f_L\simeq 1$ and $f_R\simeq 0$.  In this regime, the transport current takes the form
\begin{equation}
\frac{I}{e} \simeq
\frac{\Gamma_L \; [1+\varepsilon^2 \Gamma_i\theta/(\hbar t)^2\Gamma_R]}
{[1+\varepsilon^2 (\Gamma_L+\Gamma_i\theta)/(\hbar t)^2\Gamma_R]} ,\label{eq:RWQ}
\end{equation}
corresponding to a Lorentzian line-shape centered on the resonant condition $\varepsilon =0$.  The half-width of the peak is given by $\varepsilon_{1/2}^2\simeq (\hbar t)^2\Gamma_{Rp}/\Gamma_{Lp}$, where $\Gamma_{Lp}$ and $\Gamma_{Rp}$ correspond to barrier tunnel rates, evaluated at the peak value of the current.  Along the line from Q to R, the functions $\Gamma_L$ and $t$ remain approximately constant.  To the left of the peak, the data are nearly flat, as shown in Fig.~\ref{fig:fits-to-cuts}(b), with asymptotic behaviors determined by the bottleneck rate $\Gamma_i$.  (Note from Table~\ref{tab:table1} that $\Gamma_{iS}\lesssim \Gamma_{LS}$ along line segment RWQ.)  We conclude that $\Gamma_{iS}$ is nearly constant as a function of $\varepsilon$. The dips in the data between R and W are due to drifts in the measurement.  The suppression of the current to the left of R is probably caused by energy dependent variations of $\Gamma_i$, which are not included in our model.  It is interesting to note that the general shape of the curve described in Eq.~(\ref{eq:RWQ}) is relatively insensitive to changes in $\Gamma_R$.  This is consistent with the fact that tunneling from dot 2 to the right lead is the fastest of the tunnel rates, and it is therefore never a bottleneck.  

To the right of the resonant peak in Fig.~\ref{fig:fits-to-cuts}(b), the following conditions are satisfied:  $\theta=0$ and $-\varepsilon \gg \hbar t$, so Eq.~(\ref{eq:RWQ}) is is no longer valid.  The current in this region is more strongly suppressed than the linearized theory we discuss here predicts.

Using the fitting parameters reported in Table~\ref{tab:table1}, Eq.~(\ref{eq:full}) can be used to reconstruct the singlet and triplet bias triangles corresponding to the electron transport data shown in Figs.~\ref{fig:all-data}(g) and \ref{fig:reconstruct}(a).  The resulting theoretical fits are presented in Fig.~\ref{fig:reconstruct}(b).  The fits are quite satisfactory, and they provide strong support for the double dot theory described above.  

\subsection{Self-consistency check}
In Sec.~\ref{sec:theory}, we noted that the singlet and triplet transport channels should approximately decouple, if our theory is valid.  As a self-consistency test, we should check whether this statement is consistent with the tunneling parameters obtained from the fitting procedure.  Specifically, we want to show that the singlet density, $\rho_S=\rho_{S1}+\rho_{S2}$, is small wherever the triplet density, $\rho_T=\rho_{T1}+\rho_{T2}$, is appreciable, and vice versa.  

The formalism developed in Appendix~\ref{lindblad} and Sec.~\ref{sec:theory} allows us to compute the steady-state occupations for dots 1 and 2, as a function of the tunneling coefficients.  For either the singlet or the triplet triangles, the formalism of Sec.~\ref{sec:theory} leads to
\begin{widetext}
\begin{equation}
\rho_1+\rho_2= \frac{ \begin{array}{l}
\left\{ [4(\varepsilon/\hbar)^2+[\Gamma_L(1-f_L)+\Gamma_R(1-f_R)+\Gamma_i]^2]
[\Gamma_L\Gamma_R(f_L+f_R-2f_Lf_R)+\Gamma_i(\Gamma_Rf_R+\Gamma_Lf_L)]
\right. \\  \left. \hspace{1in}
+8t^2[\Gamma_L(1-f_L)+\Gamma_R(1-f_R)+\Gamma_i][f_L\Gamma_L+f_R\Gamma_R]
\right\}  \end{array} }
{\begin{array}{l}
\left\{ \left[ 4(\varepsilon/\hbar)^2+[\Gamma_L(1-f_L)+\Gamma_R(1-f_R)+\Gamma_i]^2 \right]
[\Gamma_L\Gamma_R(1-f_Lf_R)+\Gamma_i\Gamma_R(\theta+f_R\bar\theta)
+\Gamma_i\Gamma_L(\bar\theta +f_L\theta)] 
\right. \\  \left. \hspace{1in}
+4t^2 [\Gamma_L(1-f_L)+\Gamma_R(1-f_R)+\Gamma_i]
[\Gamma_L(1+f_L)+\Gamma_R(1+f_R)]  \right\} 
\end{array} } . \label{eq:r1r2}
\end{equation}
\end{widetext}

In Fig.~\ref{fig:reconstruct}(c), we plot $\rho_S=\rho_{1S}+\rho_{2S}$ using the fitting parameters from Table \ref{tab:table1}.  
We conclude that the singlet density does indeed vanish inside the triplet triangle, in the portions of the triangle where current flow is appreciable.
Below the singlet and triplet triangles, in the lower-right portion of Fig.~\ref{fig:reconstruct}(c), the theoretical model indicates an anomalous region of singlet occupation.  Such behavior is spurious, and it is a consequence of using energy dependent tunneling models outside their range of validity.

\subsection{Triplet relaxation and loading of excited states}
We first address the question of triplet-to-singlet relaxation.  In many experimental situations, the current is blockaded outside the singlet triangle.\cite{Johnson:2005p483,Koppens:2006p766}  We have shown that in the LET regime, the triplet channel is not necessarily blockaded.  However when the triplet loading is favored, if a (2,0) triplet decays to a (2,0) singlet faster than the singlet can unload, then current through the triplet triangle will be effectively blockaded.  We have also shown that the condition for this blockade to be lifted is that the total singlet loading rate, including loading via triplet decay, should be of the same order or smaller than the singlet unloading rate.\cite{Shaji:2008p540}  The observation of current flow in the triplet tail indicates that these conditions are met in our sample.  

In this paper, we did not explicitly consider the triplet-to-singlet decay channel.  However, the decay `current' must be bounded by the total loading current for the singlet.  As reported in Ref.~\onlinecite{Shaji:2008p540}, we can fit the resonance in Fig.~\ref{fig:fits-to-cuts}(a) close to the peak, to avoid spurious structure possibly related to cotunneling.  In this way, we obtain a bound on the triplet-singlet decay rate, given by  $\Gamma_{TS}<1.45 \times10^6 ~\text{s}^{-1}$.  This bound differs from that published previously, because it depends exponentially on the singlet-triplet splitting and the gate voltage-to-energy calibration, both of which have been determined to a greater accuracy in this paper.  The actual value of $\Gamma_{TS}$ is expected to be much smaller than this current estimate or that published previously in Ref.~\onlinecite{Shaji:2008p540}.

Finally, we can investigate selective tunneling into the triplet and singlet states.  We have shown that there is a very strong energy dependence for tunneling into the double dot.  Tunneling into a singlet state proceeds at a very different rate than tunneling into a triplet, even when they are at the same energy.  However, our analysis shows that the (2,0) triplet state is split off from the (2,0) singlet by a large amount:  $E_{ST}=173\,\mu\text{eV}$.  Electrons tunneling into these two states therefore experience very different barriers.  Based on our analysis of the transport data, we can estimate this difference by comparing the ratio of the singlet vs.\ triplet loading rates, as
\begin{equation}
\frac{\Gamma_{S,\mathrm{load}}}{\Gamma_{T,\mathrm{load}}} =
\frac{f_L(E_1)\Gamma_{LS}(E_1)}
{f_L(E_1+E_{ST})\Gamma_{LT}(E_1+E_{ST})} . \label{eq:ratio}
\end{equation}

When the triplet lies above the Fermi level of the left lead, there will be a very strong suppression of the triplet loading.  On the other hand, when both the singlet and triplet lie below the Fermi level, we observe a very strong enhancement of the triplet, as compared to the singlet.  For the tunneling parameters extracted above, this enhancement factor is on the of order 100.  However, this is actually an under-estimate, owing to our use of linearized tunnel functions.  Comparison of Figs.~\ref{fig:reconstruct}(a) and \ref{fig:reconstruct}(b) shows that, in regions with low current flow, the experimental transport current is more strongly suppressed than the theoretical prediction.  In general, the suppression of singlet tunneling in this regime should be enhanced for materials like silicon, which have relatively large effective masses.  For example, by using linearized tunneling functions and setting the Fermi functions to 1 in Eq.~(\ref{eq:ratio}), we obtain
\begin{equation}
\frac{\Gamma_{LS}(E_1)}{\Gamma_{LT}(E_1+E_{ST})} \propto
\exp \left[ E_{ST} \sqrt{2m^*W^2/(U-E_{LF})} \right] .
\end{equation}
Here, we see that the effective mass appears inside the exponential.

\begin{figure}
\includegraphics[width=3.3in]{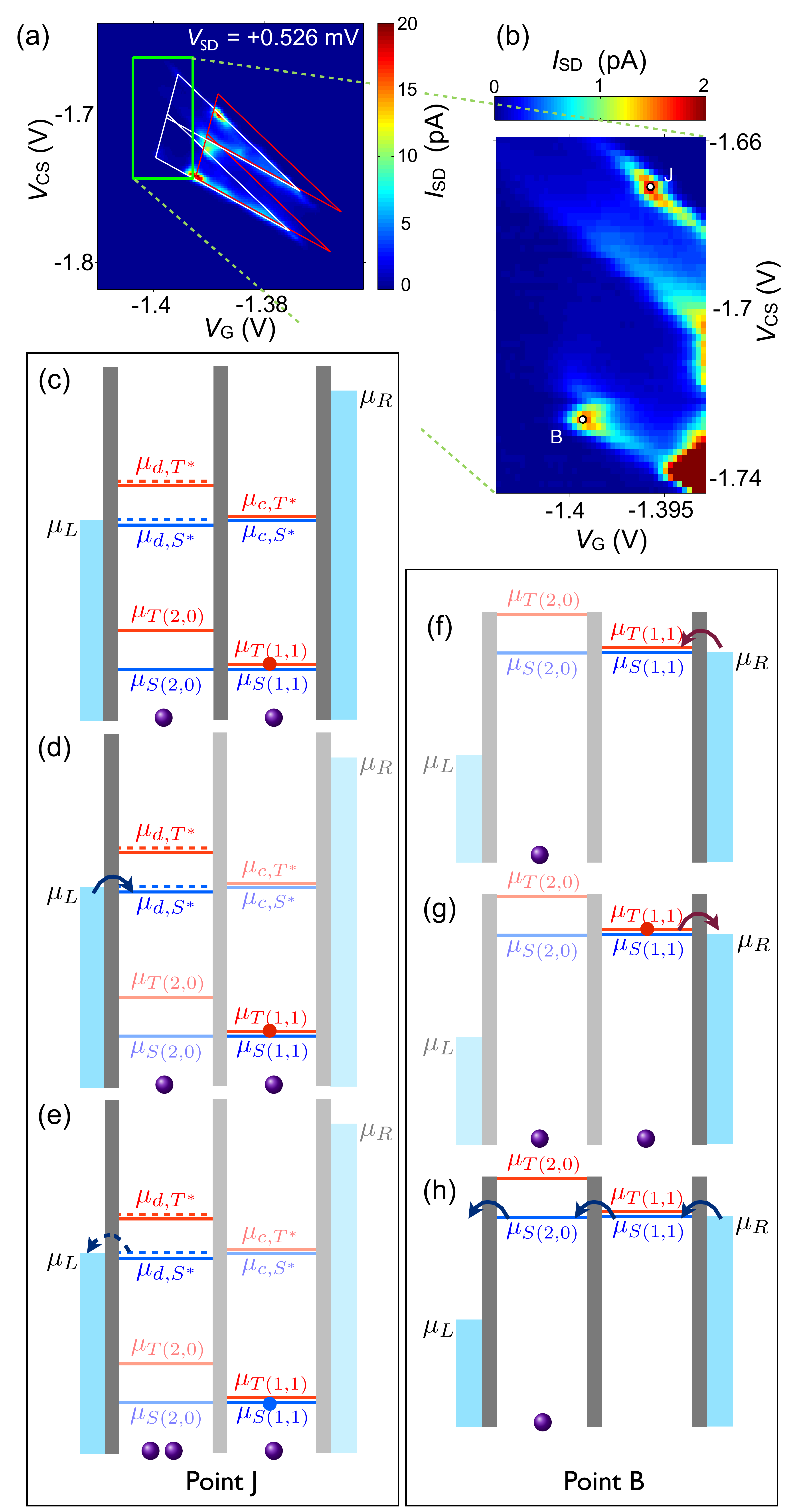}
\caption{\label{spinex}Description of the spin exchange processes between a dot and a lead at points B and J in Fig.~\ref{fig:triangle-schematic}(a).
(a) Transport data at source-drain bias of $V_{\text{SD}} = 0.526$ mV.
(b) A blow-up of the data inside the green box in panel (a). The color scale is expanded to show the current at points B and J, which are important for determining the base slope, as discussed in Sec.~\ref{sec:base}.
(c) Configuration in which the \Too state is spin blockaded, in the three electron or `hole triangle' regime.
(d) An electron can tunnel into the left dot from the left lead, forming a singlet-like (2,1) state.
(e) The singlet-like (2,1) state can emit an electron to the left lead, leaving the system in the \Soo state, and lifting spin blockade.
(f) In the `electron triangle' regime with two electrons, loading of the \Too state from the right lead results in spin blockade.
(g) The electron in the right dot can tunnel back to the right lead, allowing the right dot to be reloaded into the \Soo state.
(h) When the \Soo state is loaded from the right lead, transport can resume through the double dot. }
\end{figure}

\section{Discussion}
\label{sec:discussion}
In this paper we performed a detailed analysis of eight sets of
data measuring current through a double quantum dot.
A striking feature of the data in reverse bias is the presence of a
strong tail of current that extends outside the boundaries of the
usual bias triangle and that we attribute to lifetime-enhanced transport
(LET).
The data also contain features that are difficult to
explain using the conventional double dot transport theory, which assumes a single bias triangle.  Yet, they are explained
quite naturally when the data are fit to a pair of bias triangles, corresponding to distinct singlet and the triplet channels, as presented
in columns 2 and 4 of Fig.~\ref{fig:all-data}.
In Fig.~\ref{fig:all-data}(b), the region with strong current is broader on
the electron-triangle side than on the hole-triangle side.  The region
with strong current on the electron-triangle side lies largely within
the triplet triangle shown in red.  It is clear throughout that
tunneling through the \Too $\rightarrow$ \Ttz channel is very strong
in the reverse bias (LET) direction, and this resonance appears to
show up in the forward bias (spin blockade) direction as well.

It is worth noting that the tunnel rate between the two dots in this experiment was quite
high.  This rate was not easily tuned, because the
device was not specifically designed with a gate for this purpose.  
However, this is not a limiting factor for future experiments.
In other recent work, a double dot in Si/SiGe was specifically designed with tunable couplings, and the 
corresponding tunnel rates
were found to be highly tunable.\cite{Simmons:2009p3234} As
described in Sec.~\ref{sec:exp}, the electron occupation of our
double dot could not be absolutely determined here. However, recently, a double dot with a known
one-electron occupation has been demonstrated in a different Si/SiGe experiment.\cite{Thalakulam:2010p183104}

\textcolor{black}{The ability to control energy dependent tunneling is an important tool for measuring spin qubits.\cite{Hanson:2005p719}  Here, we observe energy dependent tunneling so strong that in many cases it deforms a bias triangle into a thin line.  Our fitting analysis indicates that the tunnel rates to the leads can change by a factor of $1/e$ when the dot chemical potential is varied by as little as 30-40~$\mu$eV.}

\textcolor{black}{The consistency of the analysis of all the data sets provides
strong evidence
that lifetime-enhanced transport occurs in a Si/SiGe double dot.
The demonstration that quantum dots can be fabricated in
Si/SiGe heterostructures that exhibit high-quality spin blockade as
well as a new transport channel that only occurs when spin relaxation
times are long is evidence that this materials system has
promise for the manufacture of devices requiring spin coherence.
}

\section*{Acknowledgements}
We thank G. Steele and I. Vink for useful discussions.
This work was supported in part by ARO and LPS
(W911NF-08-1-0482), by NSF (DMR-08325634, DMR-0805045), and by DOE
(DE-FG02-03ER46028). This research utilized NSF-supported shared
facilities at the University of Wisconsin-Madison.

\appendix

\section{Spin exchange with the leads in the spin blockade regime}
In Sec.~\ref{sec:base}, we described a method for determining the base slope by fitting a line to the points labelled B and J in Fig.~\ref{fig:triangle-schematic}.  Fig.~\ref{spinex}(b) is an expanded view of one of the spin blockade data sets, showing that points B and J are indeed clearly visible in the raw data.  At these points, and in fact along the entire segments BA and JI in Fig.~\ref{fig:triangle-schematic}, spin exchange with the leads lifts spin blockade.
In this appendix we briefly discuss this spin exchange process.  Figs.~\ref{spinex}(f)-(h) describe this spin exchange process between the right dot and the right lead at point B, while Figs.~\ref{spinex}(c)-(e)describe the spin exchange process between the left dot and the left lead at point J.

In the three electron regime at point J, the transport cycle goes from (1,1) to (2,0) to (2,1) and back to (1,1). At point J, transport is spin blockaded, because $\mu_{T(1,1)}$ lies below $\mu_{T(2,0)}$.
The set of chemical potentials labeled $\mu_{d,S^*}$, shown by the blue dashed and solid lines in Figs.~\ref{spinex}(c)-(e), refer to the energy involved in discharging the left dot from the ground, singlet-like (2,1) state to the \Soo and \Too states, respectively. At point J, when the system is blockaded in the \Too state, an electron can tunnel from the left lead to the left dot.  This corresponds to a tunneling event that is in the direction opposite to the overall electron motion under the effect of the transport bias \VSD.  Such a tunneling event can enable the formation of a singlet-like (2,1) state from the \Too state, because electrons with either spin orientation are available in the left lead.  This transition occurs at an energy given by the blue $\mu_{d,S^*}$ level (solid line).  The left dot is allowed to discharge from any one of the two $\mu_{d,S^*}$ levels. If it discharges from the higher of the two chemical potentials, the system relaxes to the singlet (1,1) state. The two $\mu_{d,S^*}$ levels are separated by the singlet-triplet splitting of the (1,1) state and are therefore very closely spaced.

\section{\label{energy-diagrams}The chemical potentials of the
  electron and hole triangles}
  
\begin{figure*}[h]
\includegraphics[width=5.6in]{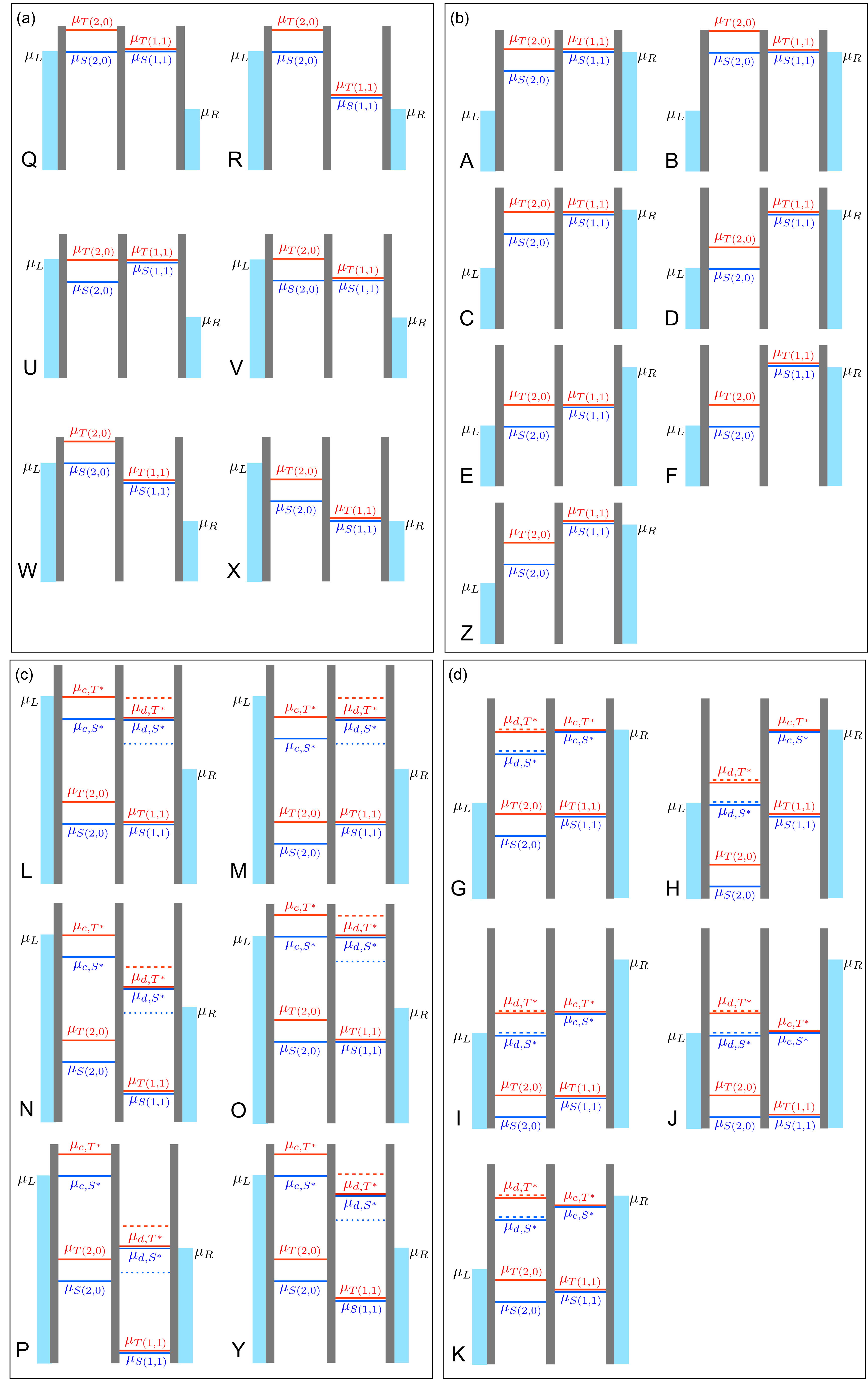}
\caption{\label{chemical-potentials}
Chemical potentials corresponding the the labelled points in Fig.~\ref{fig:triangle-schematic}. (a) Chemical potentials corresponding to the two-electron regime for reverse bias, which shows LET. (b) The two-electron regime for forward bias, which shows spin-blockade. (c) The three-electron or `hole' regime for reverse bias (LET). (d) The three-electron or `hole' regime for forward bias (spin blockade).}
\end{figure*}

Figure~\ref{chemical-potentials} shows the chemical potentials corresponding to the full set of labelled points in Fig.~\ref{fig:triangle-schematic}.

\textcolor{black}{\section{\label{lindblad}Theoretical model for quantitative
energy-dependent tunneling effects}
In this appendix we present our theoretical
treatment of energy-dependent tunneling effects that uses
the Lindblad formalism to account for both resonant and incoherent processes.}

\textcolor{black}{In our calculations, we treat the S and T transport channels independently.  Our reference state has one fixed electron in dot 1, and we consider transport that involves just three different states:  $|0\rangle$, $|1\rangle$, and $|2\rangle$.
Here, $|0\rangle$ refers to the state with no additional electrons, $|1\rangle$ refers to the state with one additional electron on dot 1 [the (2,0) charge configuration], and $|2\rangle$ refers to the state with one additional electron on dot 2 [the (1,1) charge configuration].  Coherent evolution is controlled by the Hamiltonian
\begin{equation}
H =\frac{1}{2}\varepsilon (|1\rangle \langle 1| - |2\rangle \langle 2| ) 
+\hbar t(|1\rangle \langle 2|+|2\rangle \langle 1|) ,
\end{equation}
where $t$ is the tunnel coupling between the two dots.}

\textcolor{black}{We now couple this system
to the environment,
using the Lindblad formalism.\cite{PreskillNotes,NielsenBook}
Tunneling from the left lead to dot~1 is described by the Lindblad operator
\begin{equation}
L_{L1}=\sqrt{f_L(E_1)\Gamma_L(E_1)} \: |1\rangle \langle 0| ,
\end{equation}
where $f_L(E_1)=f(E_1-E_{LF})$ is the Fermi function of the left lead, and $\Gamma_L$ is the tunnel coupling between the lead $L$ and dot~1.  Both $f_L$ and $\Gamma_L$ depend on energy, but, for brevity, we will suppress the energy dependence in the notation.  The other relevant Lindblad operators are given by
\begin{eqnarray}
L_{1L} &=& \sqrt{(1-f_L)\Gamma_L} \: |0\rangle \langle 1| , \\
L_{2R} &=& \sqrt{(1-f_R)\Gamma_R} \: |0\rangle \langle 2| , \\
L_{R2} &=& \sqrt{f_R\Gamma_R} \: |2\rangle \langle 0| , \\
L_{12} &=& \sqrt{\theta \Gamma_i} \: |2\rangle \langle 1| , \label{eq:L12} \\
L_{21} &=& \sqrt{\bar\theta \Gamma_i} \: |1\rangle \langle 2| . \label{eq:L21}
\end{eqnarray} 
Note that reverse processes (from right to left), such as $L_{1L} $, are also included here.  The latter play a role along the edges of the bias triangle.  For example, an electron may enter dot 1 from lead $L$ and then exit back to lead $L$.  Such processes do not directly affect the steady-state current, but they do affect the current indirectly, because, while the dot is occupied, it cannot be occupied by a second, right-moving electron.}

\textcolor{black}{In Eqs.~(\ref{eq:L12}) and (\ref{eq:L21}), the incoherent tunneling between the two dots involves phonon emission or absorption processes.  We have accounted for these phonon effects through the $\theta$-functions.  At high temperatures, the $\theta$-functions may possess considerable structure.  However for low temperature applications, we assume that
$\theta =\Theta(\varepsilon)$ and $\bar\theta =\Theta(-\varepsilon)$, where the step function $\Theta(\varepsilon)$ takes the values 0, when $\varepsilon<0$, 1/2  when $\varepsilon=0$, and 1 when $\varepsilon>0$.  More general forms for $\theta$ can be substituted, as appropriate.  We also note that the inelastic, interdot tunnel coupling $\Gamma_i$ is a weak, even function of $\varepsilon$.  Below, we show that our data are consistent with $\Gamma_i=(\text{constant})$ throughout most of the bias triangle.}

\textcolor{black}{The evolution of the density operator is described by the rate equation
\begin{equation}
\dot{\hat\rho} = -\frac{i}{\hbar}[H,{\hat\rho}] +\sum_{j}\left[ L_j{\hat\rho}_j^\dagger - \frac{1}{2}
\left\{{\hat\rho} ,L_j^\dagger L_j \right\} \right] , \label{eq:Lindblad}
\end{equation}
where $\hat\rho$ is the density matrix.
In general, $\hat\rho$ must satisfy the normalization condition
$1=\rho_0+\sum_k (\rho_{1k}+\rho_{2k} )$, where the diagonal terms $\rho_0$, $\rho_{1k}$, and $\rho_{2k}$ describe the probability of being in a given occupation state, and the sum over $k$ includes the singlet and three triplet channels.  In the absence of any decay processes between the triplet and singlet states, this normalization condition is the only coupling between the singlet and triplet sectors, since it ensures that a triplet cannot be formed when a singlet state is occupied, and vice versa.  However, in the LET regime, we have shown that the $(2,0)$ singlet state unloads much faster than it loads, and that the unloading of the (1,1) singlet is similarly fast (or faster).\cite{Shaji:2008p540}   These conditions are equivalent to the statement that $\rho_{1S}+\rho_{2S} \ll 1$ wherever $\rho_{1T}+\rho_{2T}$ is appreciable, and vice versa.  Since LET behavior is observed in our samples, we make a singlet-triplet decoupling approximation, such that the normalization
\begin{equation}
1\simeq \rho_0+\rho_1+\rho_2  ,\label{eq:rhonorm}
\end{equation}
applies to both the singlet and triplet channels.  The resulting rate equations are correspondingly simplified.  }

\textcolor{black}{Since there is no coherent coupling between state $|0\rangle$ and states $|1\rangle$ and $|2\rangle$, the density operator for a single channel can be defined as 
\begin{equation}
\hat\rho = \rho_0 |0\rangle \langle 0| + \rho_1 |1\rangle \langle 1| +
 \rho_2 |2\rangle \langle 2| +  \rho_{12} |1\rangle \langle 2| +  \rho_{21} |2\rangle \langle 1| .
 \end{equation}
We may then use Eq.~(\ref{eq:rhonorm}) to eliminate $\rho_0$ from the rate equations defined in Eq.~(\ref{eq:Lindblad}).  
Steady-state solutions are obtained by requiring that $\dot{\hat\rho} =0$.  }

\textcolor{black}{The current operator is defined as
\begin{equation}
\hat{I}/e=it(|1\rangle \langle 2|-|2\rangle \langle 1|)
+\Gamma_i(\theta |1\rangle \langle 1|-\bar\theta |2\rangle \langle 2|) ,
\end{equation}
and it involves both coherent and incoherent components.
The steady-state current is given by
 $I = \text{Tr} (\rho \hat{I})$.  Using the steady-state rate equations, the result can be expressed in terms of density coefficients:  
\begin{equation}
I/e =(1-f_R)\Gamma_R\rho_2-f_R\Gamma_R\rho_0 . \label{eq:Irho}
\end{equation}
In this form, the current is simply expressed as the net tunneling rate between dot 2 and lead $R$.  }

\textcolor{black}{By solving for the density coefficients, Eq.~(\ref{eq:Irho}) can be expressed entirely in terms of tunneling rates and Fermi functions, yielding the following result for single-channel transport in a two-electron double dot:
\begin{widetext}
\begin{equation}
\frac{I}{e}= \Gamma_L\Gamma_R \frac{ \begin{array}{l}
\left\{ 4t^2(f_L-f_R)[\Gamma_L(1-f_L)+\Gamma_R(1-f_R)+\Gamma_i] 
\right. \\ \left. \hspace{1in}
+ \Gamma_i\left( 4(\varepsilon/\hbar)^2+[\Gamma_L(1-f_L)+\Gamma_R(1-f_R)+\Gamma_i] ^2 \right)
[f_L(1-f_R)\theta -f_R(1-f_L)\bar\theta ] \right\} \end{array} }
{\begin{array}{l}
\left\{ \left[ 4(\varepsilon/\hbar)^2+[\Gamma_L(1-f_L)+\Gamma_R(1-f_R)+\Gamma_i]^2 \right]
[\Gamma_L\Gamma_R(1-f_Lf_R)+\Gamma_i\Gamma_R(\theta+f_R\bar\theta)
+\Gamma_i\Gamma_L(\bar\theta +f_L\theta)] 
\right. \\  \left. \hspace{1in}
+4t^2 [\Gamma_L(1-f_L)+\Gamma_R(1-f_R)+\Gamma_i]
[\Gamma_L(1+f_L)+\Gamma_R(1+f_R)]  \right\} 
\end{array} } . \label{eq:full2app}
\end{equation}
\end{widetext}
When the singlet and triplet channels are fully decoupled, as we assume here, the total current is expressed as a sum of terms like Eq.~(\ref{eq:full2app}), one for each channel.  }

\textcolor{black}{As an initial check on our result, we consider a known limit.  For the case of pure coherent tunneling between dots 1 and 2, we take the limit $\Gamma_i\rightarrow 0$.  In the interior of the bias triangle where $f_L\simeq 1$ and $f_R\simeq 0$, Eq.~(\ref{eq:full2app}) immediately reproduces the resonant tunneling results obtained in Refs.~\onlinecite{Nazarov:1993p57} and \onlinecite{Stoof:1996p1050}. }


\end{document}